\documentclass[twocolumn]{aastex631}

\usepackage{amsmath, amssymb, amsfonts}
\usepackage{mathtools}
\usepackage{bm}
\usepackage{url}
\usepackage{cancel}
\usepackage{epstopdf}
\usepackage{subfigure}
\usepackage{booktabs}

\begin{document}

\title{X-rays from Low-mass Active Galaxies in the Galaxy And Mass Assembly (GAMA) Survey}

\author{Stellan Bechtold}
\affil{Department of Physics, Montana State University, Bozeman, MT 59717, USA}

\author{Amy Reines}
\affil{Department of Physics, Montana State University, Bozeman, MT 59717, USA}

\author{Richard M. Plotkin}
\affil{Physics Department, University of Nevada, Reno, 1664 N. Virginia Street, Reno NV 89557, USA}
\affil{Nevada Center for Astrophysics, University of Nevada, Las Vegas, NV 89154, USA}

\author{Sheyda Salehirad}
\affil{Department of Physics, Montana State University, Bozeman, MT 59717, USA}

\begin{abstract}

We present new \textit{Chandra} observations of seven broad-line AGNs in low-mass galaxies ($M_\star < 5 \times 10^9\,M_\odot$) selected from the GAMA sample of \citet{salehirad_hundreds_2022}. X-ray emission is detected in six galaxies, with luminosities spanning $\log(L_{\rm 0.5-8\,keV}/{\rm erg\,s^{-1}})\approx40.5$--43.8. Their X-ray luminosities generally exceed predictions for X-ray binary populations, and their proximity to the optical galaxy nuclei favors accreting massive black holes as the dominant emission source. Spectral fits for four sources yield photon indices $\Gamma\approx1.7$--2.2, with little evidence for intrinsic absorption in most cases. Virial black hole masses and hard X-ray bolometric corrections imply nominal Eddington ratios $\lambda_{\rm Edd}\approx0.004$--1.1, although uncertainties are substantial. Optical tracers generally imply higher bolometric luminosities and Eddington ratios, particularly for the X-ray-undetected galaxy. The X-ray-bright sources are broadly consistent with optical AGN scaling relations involving H$\alpha$ and [O~\textsc{iii}], whereas most lie below the established X-ray--mid-infrared relation, potentially reflecting host-galaxy contamination. Four sources are consistent with the expected UV-to-X-ray spectral slope relation, while three show evidence for relative X-ray weakness. These results reveal diverse X-ray properties among broad-line AGNs in low-mass galaxies and highlight the importance of multiwavelength diagnostics for understanding black hole accretion and interpreting broad-line AGN candidates at high redshift.

\end{abstract}

\section{Introduction} \label{sec:intro}

Practically all massive galaxies host supermassive black holes (SMBHs) at their centers \citep{kormendy_coevolution_2013}. However, the mechanisms by which primeval SMBH "seeds" are formed and how they grow to become the massive SMBHs we see in modern galaxies are not well understood \citep{greene_intermediate-mass_2020}.  We approach the issue of constraining seeding models by studying AGN hosted in dwarf galaxies ($M_* \leq 3 \times 10^9 M_{\odot}$) in the nearby universe (see  \citealt{reines_hunting_2022} for a review). Compared to massive galaxies, dwarf galaxies have relatively clean merger histories \citep{bellovary_first_2011}, making their central BHs with typical masses $M_{\rm BH} \sim 10^4-10^6~M_\odot$ better representations of initial BH seeds.

We are able to search for dwarf galaxy central BHs in the form of active galactic nuclei (AGN) and characterize the emission from AGN to constrain their properties. X-ray observations provide one of the most effective ways to identify AGN candidates in dwarf galaxies \citep[e.g.,][]{lemons_x-ray_2015, pardo_x-ray_2016, birchall_x-ray_2020, bykov_srgerosita_2024, sacchi_x-ray_2024, eberhard_new_2025}. Emission in the X-ray band originates from the innermost regions of the accretion flow, offering direct insight into black hole accretion processes. Moreover, AGNs often exhibit X-ray luminosities that exceed the expected contribution from normal stellar-mass X-ray binaries \citep[XRBs;][]{lehmer_chandra_2010, mineo_x-ray_2012, lehmer_empirical_2024}, enabling a clear distinction between accretion onto a central black hole and stellar processes. However, ultra-luminous X-ray sources (ULXs) can reach luminosities comparable to low-luminosity AGN and therefore represent a potential source of contamination, particularly in star-forming galaxies \citep{feng_ultraluminous_2011,kaaret_ultraluminous_2017}.

X-ray observations also provide the most robust means of confirming AGNs in dwarf galaxies selected via methods at other wavelengths \citep[e.g.,][]{reines_deep_2016, baldassare_x-ray_2017, latimer_chandra_2021, kimbrell_luminous_2024}. Optical spectroscopic diagnostics have been used to identify large samples of AGN candidates in dwarf galaxies for more than a decade \citep{reines_dwarf_2013, moran_black_2014, pucha_tripling_2025}. Narrow-line diagnostic diagrams (e.g., BPT; \citealt{baldwin_classification_1981}) use emission line ratios to classify the dominant gas ionization mechanism in galaxies \citep{kewley_theoretical_2001, kauffmann_host_2003, kewley_host_2006}, while broad H$\alpha$ emission can signify dense gas orbiting within the broad-line region just light-days from the central BH. For these broad-line AGNs, it is possible to estimate the mass of the BH \citep{greene_estimating_2005,reines_dwarf_2013}. The velocity of the gas in the broad line region is inferred from the width of the broad H$\alpha$ emission line, while the size of the broad line region is estimated based on the luminosity of the broad emission \citep{kaspi_reverberation_2000, peterson_central_2004, greene_estimating_2005, bentz_lick_2009, bentz_low-luminosity_2013}.
We note that broad Balmer emission is not uniquely associated with AGN activity and can also arise from transient phenomena such as Type II supernovae or extreme stellar processes \citep{reines_dwarf_2013}. However, the presence of AGN-like narrow-line ratios in addition to broad emission strongly supports an AGN origin \citep{baldassare_multi-epoch_2016}.

\citet{salehirad_hundreds_2022} identified 47 low-mass galaxies in the Galaxy and Mass Assembly (GAMA) Survey with narrow- and broad-emission-line signatures indicating the presence of an AGN. Here, we present \textit{Chandra X-ray Observatory} observations of seven of these objects with reported stellar masses $M_\star < 5 \times 10^9~M_\odot$ and BH masses $M_{\rm BH} \sim 3 \times 10^5 - 3 \times 10^7~M_\odot$. While the sample size is modest, systems with both virial BH mass estimates and sensitive X-ray observations remain relatively rare, particularly among broad-line AGNs hosted by low-mass galaxies \citep{baldassare_x-ray_2017, chilingarian_population_2018}. This sample therefore provides an important addition to the growing census of accreting low-mass BHs and enables a uniform investigation of their X-ray spectral and accretion properties.

The structure of this paper is as follows. In Section \ref{sec:observations}, we describe the sample selection. Section \ref{sec:chandra_observations} introduces the \textit{Chandra} observations, data reduction procedures, and astrometric analysis. Section \ref{sec:analysis} presents the X-ray analysis and results, including photometry, spectral modeling, bolometric luminosities and Eddington ratios, estimates of the expected X-ray emission from stellar processes, and comparisons with multiwavelength AGN scaling relations. In Section \ref{sec:discussion}, we assess the origin of the X-ray emission and discuss the multiwavelength properties of the sources, including evidence for possible X-ray weakness. Finally, Section \ref{sec:conclusion} summarizes our conclusions. Throughout this paper, we adopt a flat $\Lambda$CDM cosmology with $H_0 = 70~\mathrm{km~s^{-1}~Mpc^{-1}}$ and $\Omega_{\rm m}=0.3$.

\begin{figure}
    \centering
    \includegraphics[width=\linewidth]{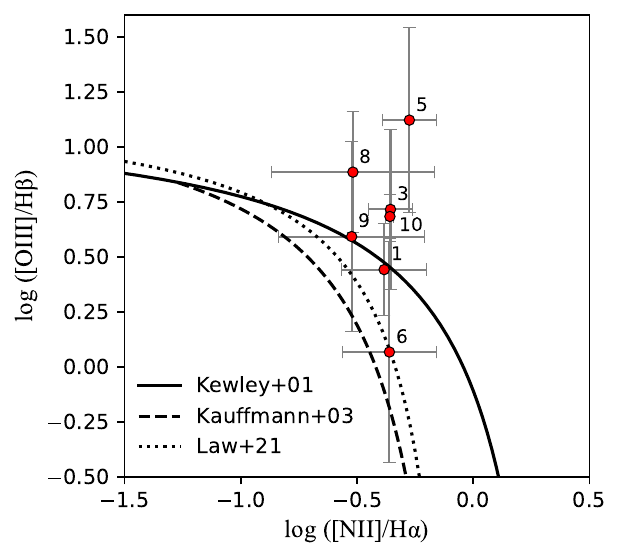}
    \caption{Target galaxy positions on the [N \textsc{ii}]-BPT diagram 
    \citep{baldwin_classification_1981}. The \citet{kauffmann_host_2003} 
    and \citet{kewley_theoretical_2001} demarcation lines are shown 
    in \citet{kewley_host_2006}. The dotted curve shows the 
    $1\sigma$ cold-gas demarcation from \citet{law_sdss-iv_2021}. 
    Errors are shown at the 95\% confidence level.}
    \label{fig:NII_BPT}
\end{figure}

\begin{figure}
    \centering
    \includegraphics[width=\linewidth]{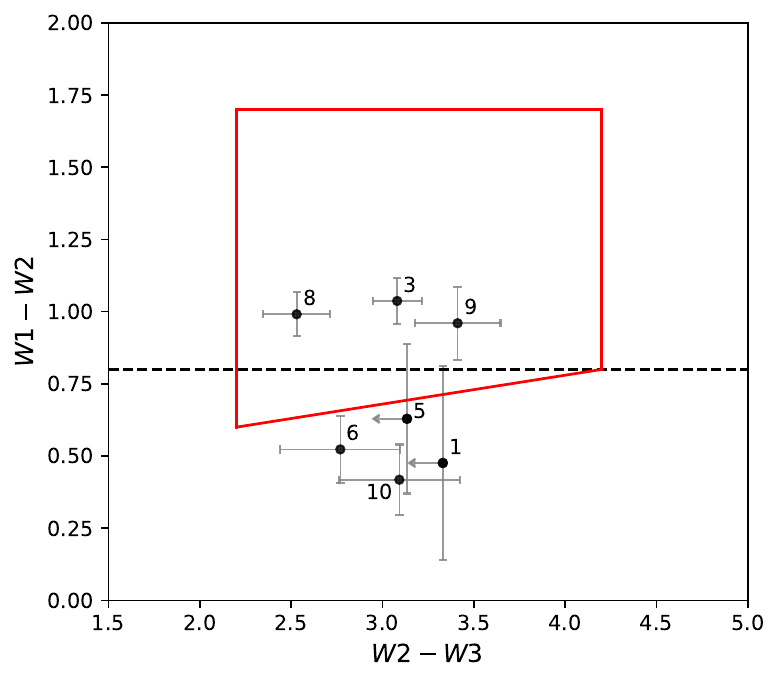}
    \caption{Our target galaxies on the WISE W1--W2 color vs. W2--W3 color plot. Error bars represent $95\%$ uncertainties. The arrows for IDs 1 and 5 indicate W2--W3 colors calculated using the 95\% upper limits in W3 and the measured W2 magnitudes. The region enclosed by red lines is the AGN selection box from \citet{jarrett_spitzer-wise_2011}. The dashed black line denotes the AGN demarcation line W1--W2 = 0.8 from \citet{stern_mid-infrared_2012}.}
    \label{fig:WISE}
\end{figure}

\begin{deluxetable*}{cccccccccc}
\tabletypesize{\footnotesize}
\tablecaption{\label{tab:chandra_obs}\textit{Chandra} Observations}
\tablehead{
\colhead{ID} &
\colhead{CATAID} &
\colhead{RA} &
\colhead{Dec} &
\colhead{$\log M_*/M_\odot$} &
\colhead{z} &
\colhead{BPT} &
\colhead{Obs. Date} &
\colhead{Obs. ID} &
\colhead{Exp. (ks)}
}
\colnumbers
\startdata
1 & 718498 & 180.52022808 & 1.83566996 & 9.23 & 0.13933 & Comp. & 2025-07-30 & 30009 & 14.65\\
3 & 5240292 & 343.2688411 & -33.9772443 & 9.59 & 0.2237 & AGN & 2025-08-22 & 30010 & 8.96\\
5 & 2258819 & 30.874079 & -9.491468 & 9.60 & 0.2517 & AGN & 2026-01-17 & 30011 & 18.83\\
6 & 593724 & 217.55753021 & -0.17237251 & 9.61 & 0.10384 & Comp. & 2025-05-17 & 30012 & 16.86\\
8 & 5275222 & 345.3551208 & -34.2113545 & 9.64 & 0.21019 & AGN & 2025-04-04 & 30013 & 5.74\\
9 & 484908 & 213.24339724 & -1.80648722 & 9.64 & 0.11928 & AGN & 2025-06-14 & 30014 & 15.87\\
10 & 238411 & 214.24744918 & 1.57398776 & 9.65 & 0.0561 & AGN & 2025-08-16 & 30015 & 20.80\\
\enddata
\tablecomments{Columns from left to right: (1) Galaxy/Source ID assigned in this paper, (2) GAMA catalog ID (CATAID), (3) object Right Ascension in degrees (RA), (4) object Declination in degrees (Dec.), (5) galactic stellar mass in solar masses on the logarithmic scale ($\log M_*/M_{\odot}$), (6) spectroscopic redshift (z), (7) [NII]-BPT location, (8) \textit{Chandra} observation date, (9) \textit{Chandra} observation ID, (10) \textit{Chandra} observation effective exposure time.}
\end{deluxetable*}

\section{Sample} \label{sec:observations}

The low-mass galaxies that are the subjects of this work are taken from the AGN candidate sample assembled via a suite of diagnostics in \citet{salehirad_hundreds_2022}. In their work, \citet{salehirad_hundreds_2022} used optical spectroscopy from the Galaxy And Mass Assembly (GAMA) Survey DR4 \citep{driver_galaxy_2022} to identify a sample of active galactic nuclei (AGN) candidates in low-mass galaxies with stellar masses $\log (M_*/M_{\odot}) \leq 10$. From this parent sample, 47 galaxies were identified that exhibit broad H$\alpha$ emission and are classified as AGN candidates based on the narrow emission-line [NII]-BPT diagnostic diagram. These galaxies lie either in the AGN region or the Composite region of the diagram, with 25 objects classified as AGN and 22 as Composite. Galaxies in the Composite region show evidence that both star formation and AGN activity contribute to their narrow emission-line fluxes.

We obtained new observations of seven galaxies from this subsample using the \textit{Chandra X-ray Observatory}. These galaxies were selected for follow-up because they represent some of the lowest stellar masses in the broad-line sample and are sufficiently bright to provide a reasonable chance of X-ray detection with moderate exposure times. As shown in Fig. \ref{fig:NII_BPT}, five of the galaxies lie in the AGN region of the [N \textsc{ii}]-BPT diagram, while the remaining two lie in the composite region.

\begin{figure*}
\centering
\includegraphics[width=0.24\textwidth]{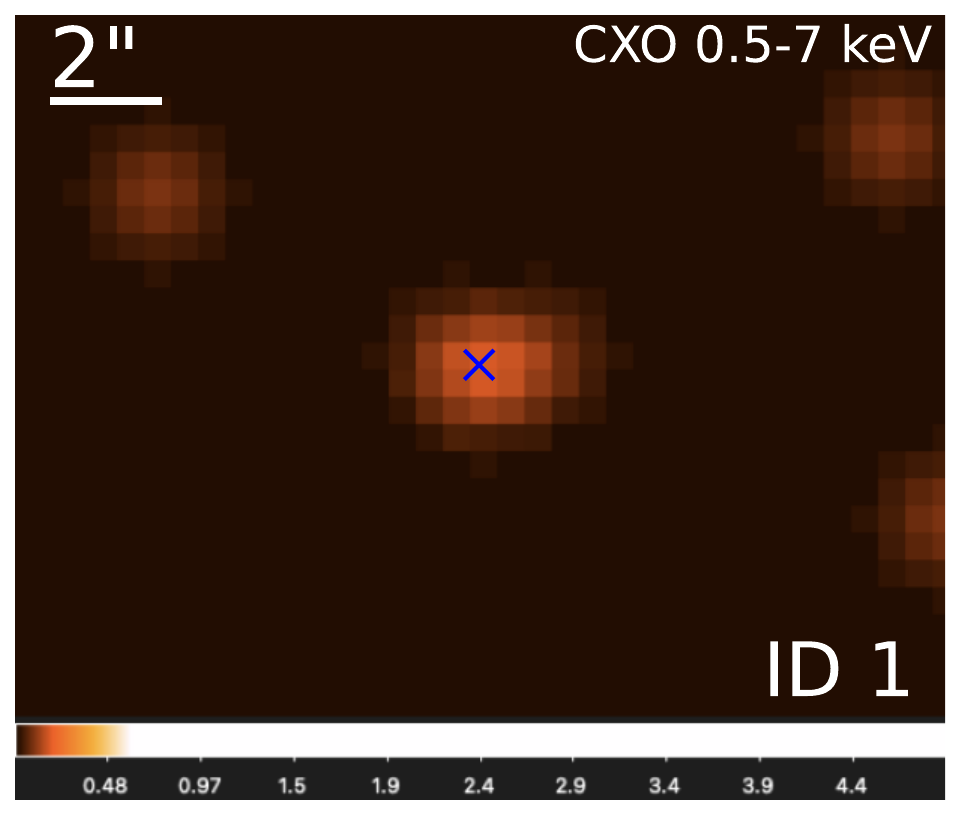}
\includegraphics[width=0.24\textwidth]{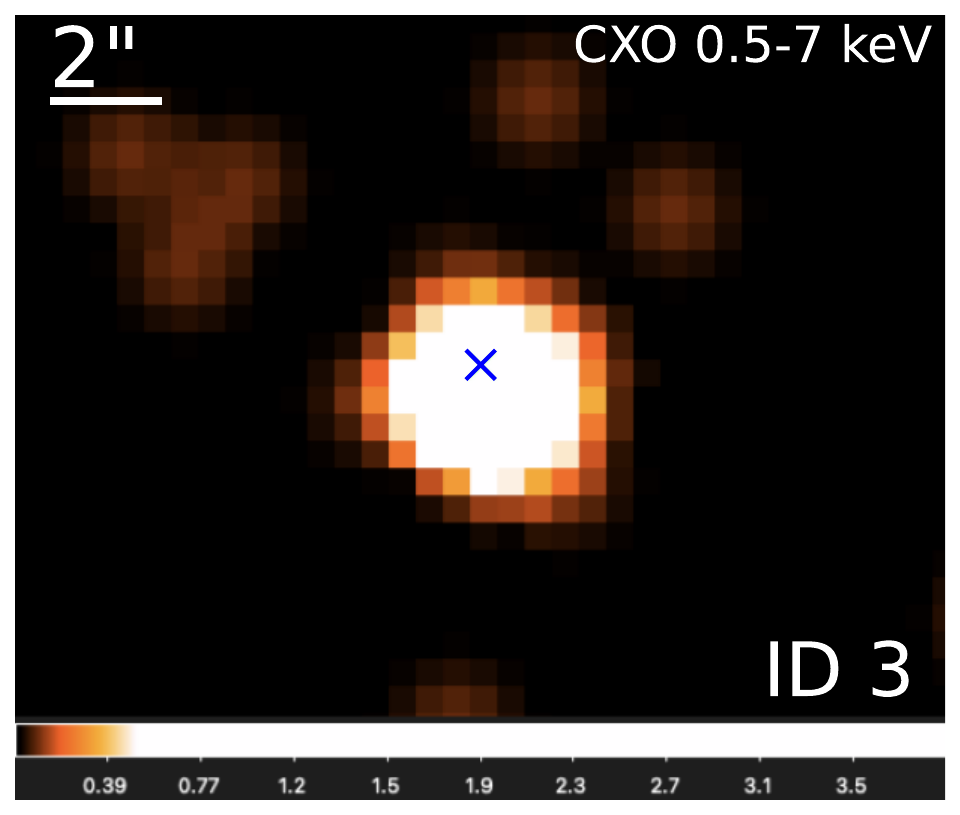}
\includegraphics[width=0.24\textwidth]{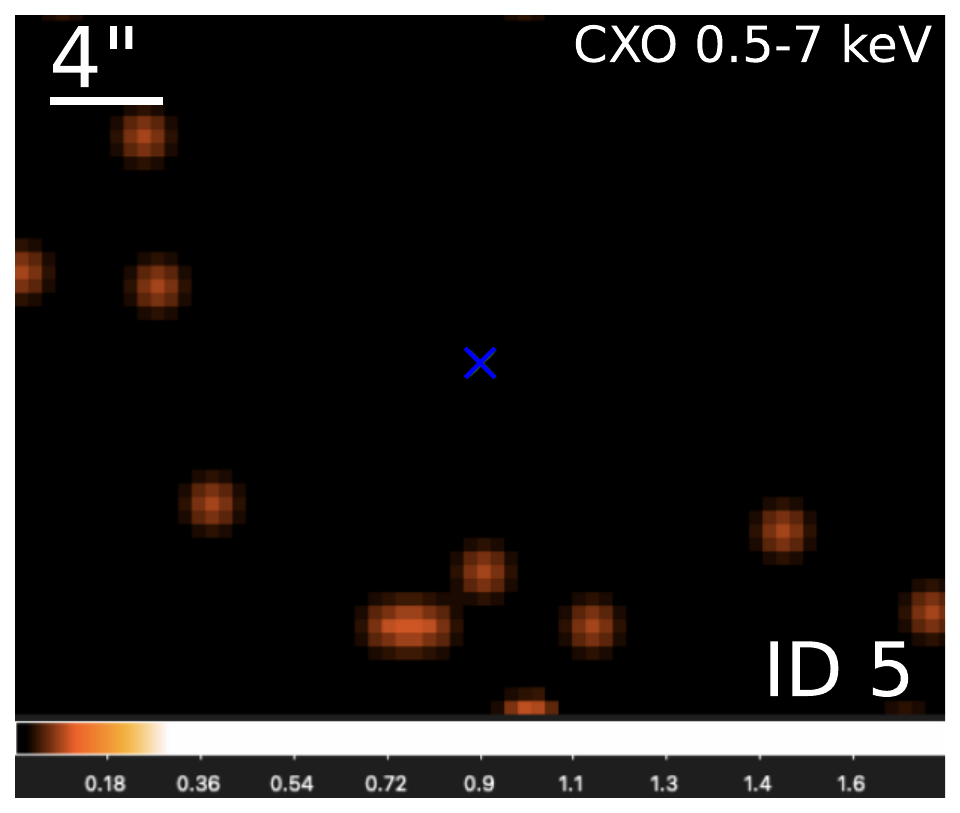}
\includegraphics[width=0.24\textwidth]{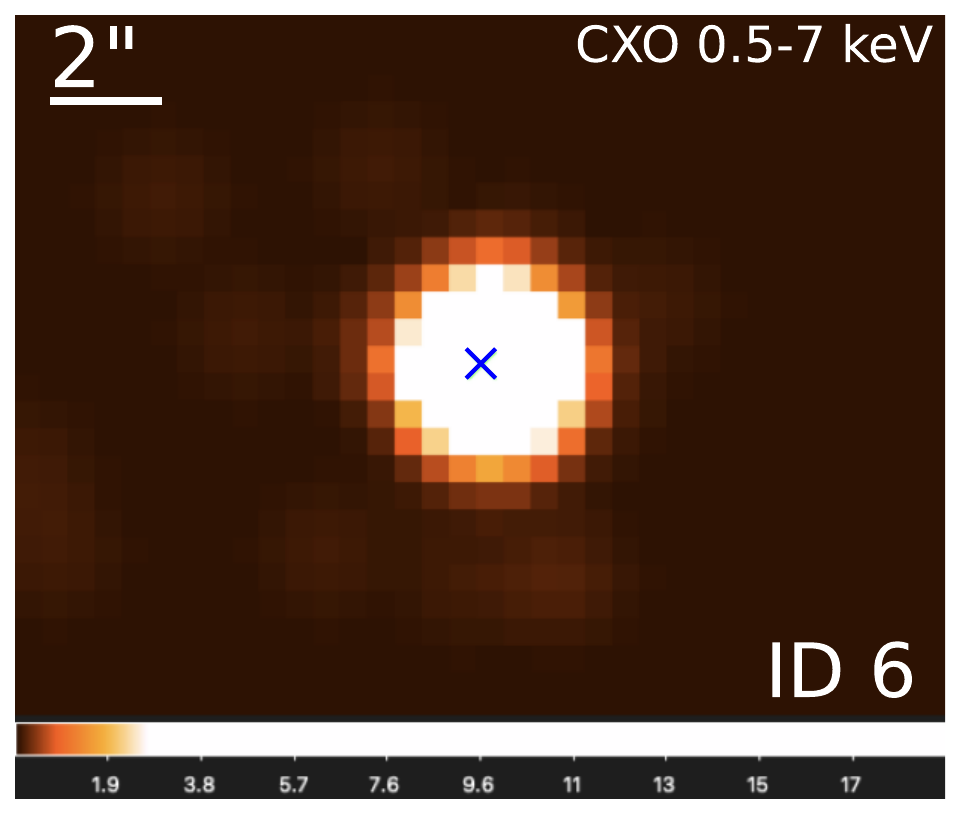}
\includegraphics[width=0.24\textwidth]{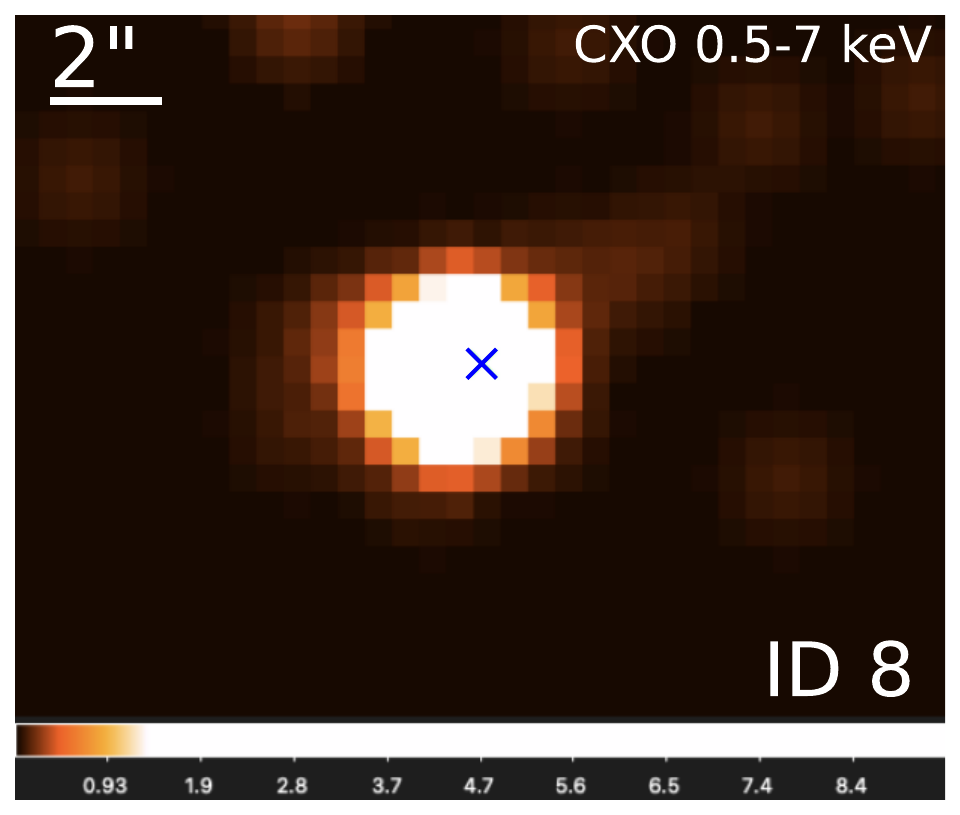}
\includegraphics[width=0.24\textwidth]{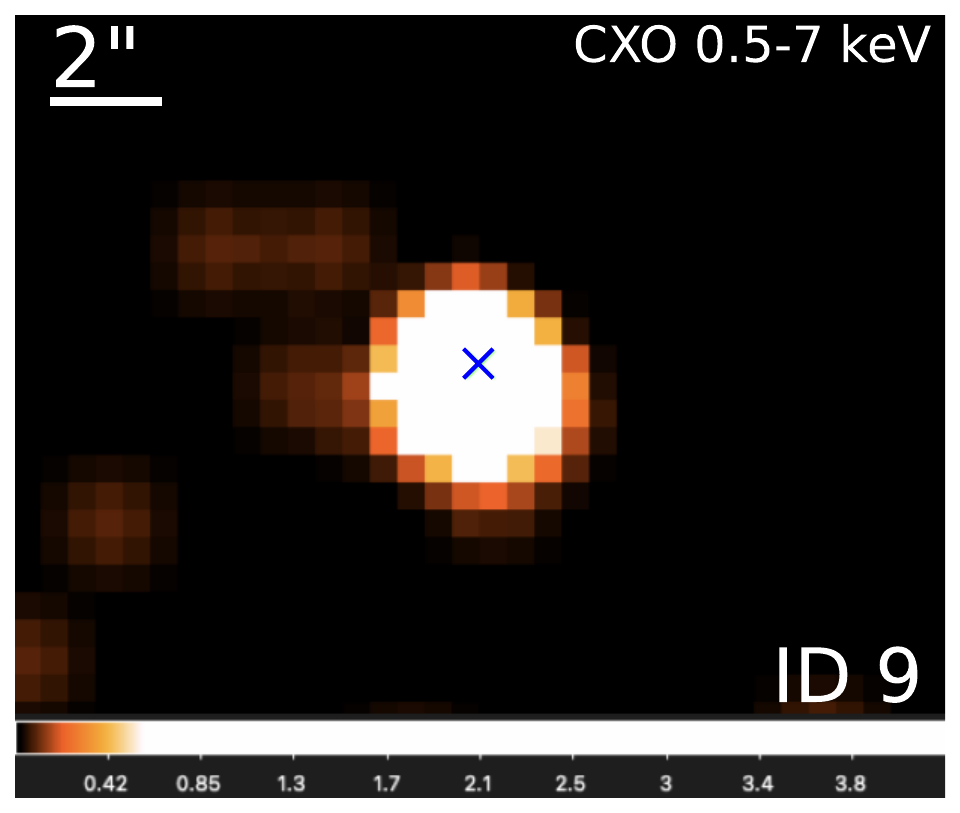}
\includegraphics[width=0.24\textwidth]{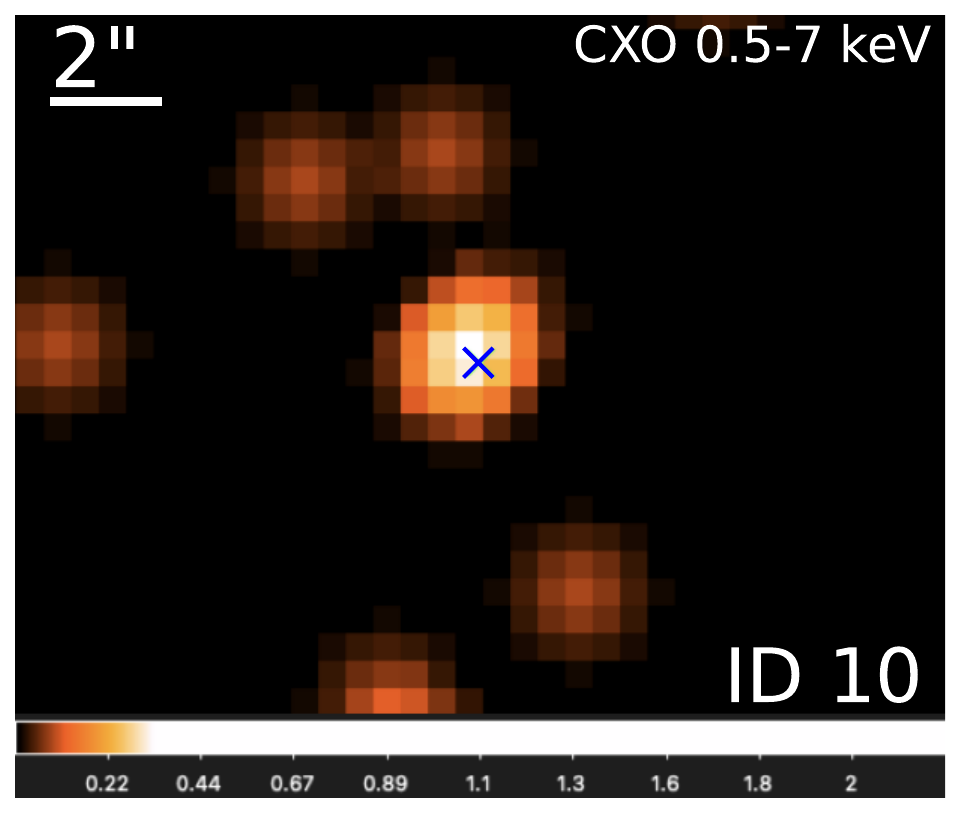}

\caption{0.5--7 keV \textit{Chandra X-ray Observatory} images of each target
galaxy, smoothed with a 3-pixel-radius Gaussian kernel. The GAMA optical
center of each galaxy is marked with a blue X.}
\label{fig:cutouts}
\end{figure*}

We additionally compare the updated [N \textsc{ii}]-BPT classifications of \citet{law_sdss-iv_2021}. Six of the seven sources lie above the \citet{law_sdss-iv_2021} cold-gas demarcation, while ID 6 falls just within the star-forming region. Therefore, the [N \textsc{ii}]-BPT classification of ID 6 is sensitive to the adopted diagnostic boundary, weakening the narrow-line evidence for AGN activity in this source. In addition to broad H$\alpha$ emission and narrow-line ratios indicating an AGN, ID 8  also has a [Ne \textsc{v}] detection in its optical spectrum  \citep{salehirad_hundreds_2022}. Moreover, three galaxies in our sample  (IDs 3, 8, and 9) have mid-IR colors indicating dust heating by an AGN  (Fig. \ref{fig:WISE}), providing additional support for the presence of  AGNs in these galaxies. The remaining four galaxies lie outside the AGN  selection region in the WISE color--color plot, which may result from dilution of the AGN emission by their host galaxies. Although broad Balmer emission is not uniquely associated with AGN activity and can also arise from transient phenomena such as Type II supernovae or extreme stellar processes \citep{reines_dwarf_2013, baldassare_multi-epoch_2016}, the combination of broad H$\alpha$ emission and narrow-line diagnostics supports these galaxies as likely AGN hosts, although the strength of narrow-line classifications varies across the sample.The [Ne~\textsc{v}] detection in ID 8 and the mid-IR colors of IDs 3, 8, and 9 provide additional evidence for AGN activity in individual sources.

For convenience, we assign each galaxy a unique ID corresponding to its relative stellar mass rank within the original sample of 47 broad-line galaxies, ordered from lowest to highest mass. Thus, ID 1 corresponds to the lowest stellar-mass galaxy in the parent broad-line sample.
The seven galaxies have stellar masses in the range $9.23 \leq \log (M_*/M_{\odot}) \leq 9.65$ and redshifts between $0.0561 \leq z \leq 0.2517$. Stellar masses and spectroscopic redshifts are provided by the GAMA survey (Table \ref{tab:chandra_obs}). Black hole mass estimates were derived using standard virial techniques applied to the broad H$\alpha$ emission. \citet{salehirad_hundreds_2022} report black hole masses spanning $5.5 \leq \log (M_{\rm BH}/M_{\odot}) \leq 7.5$ for the galaxies in our sample (see \S\ref{sec:accretion}).

\begin{figure*}
    \centering
    \includegraphics[width=0.9\textwidth]{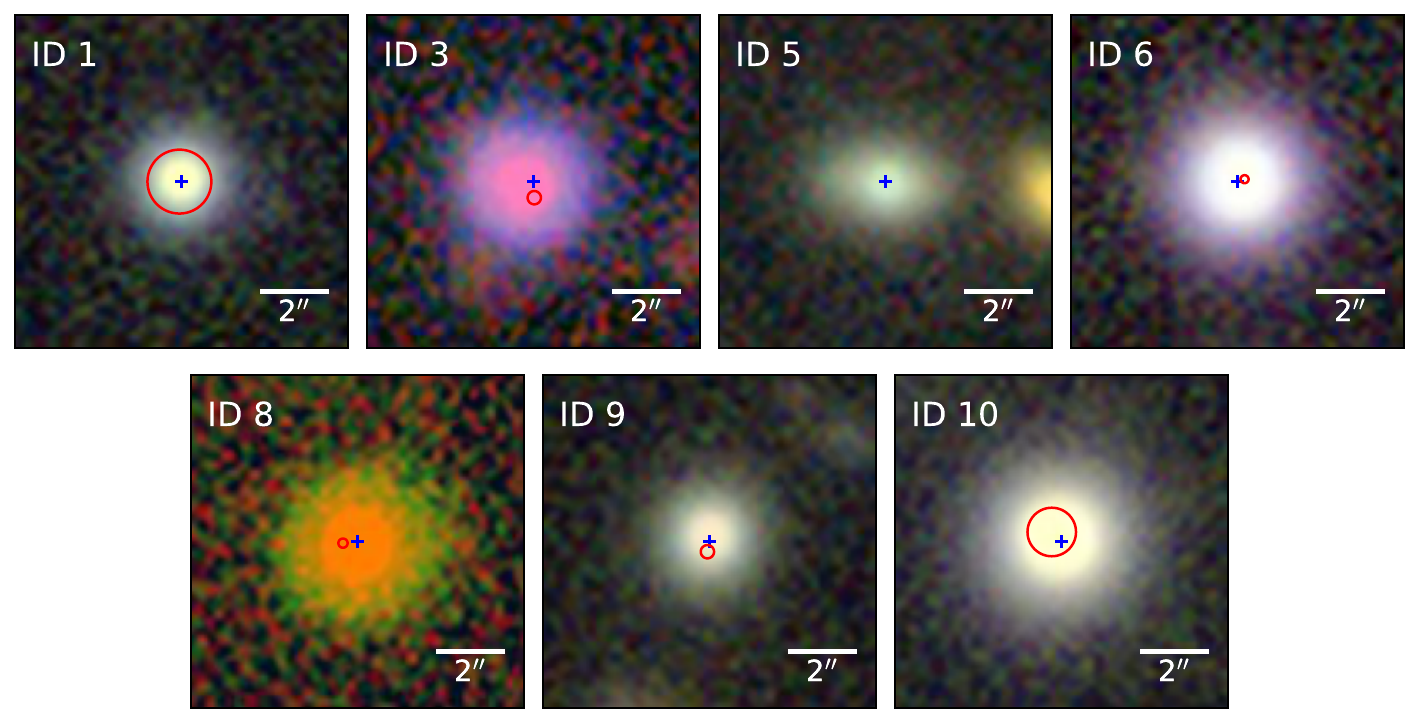}
    \caption{X-ray positions and uncertainties overlaid on \textit{grz} DESI Legacy Imaging Survey SkyViewer images of each target galaxy \citep{xu_desi_2023}. Blue crosses mark the GAMA coordinates of the optical centers, while red circles are centered on the X-ray source positions, with radii corresponding to the 95\% positional uncertainties calculated using Eq. 12 of \citet{kim_chandra_2007}. For ID 1, the circle is instead centered on the GAMA optical coordinates because no independent X-ray source position was identified with \texttt{wavdetect} (see \S\ref{sec:astrometry}). No X-ray position or positional uncertainty is shown for ID 5, which was not detected in the X-ray observations.}
    \label{fig:xray_positions}
\end{figure*}

\section{Chandra X-ray Observations} \label{sec:chandra_observations}

This paper employs \textit{Chandra} datasets obtained by the \textit{Chandra X-ray Observatory} and contained in \dataset[DOI: 10.25574/cdc.718]{https://doi.org/10.25574/cdc.718}. Observations of our targets were taken with the Advanced CCD Imaging Spectrometer S-array (ACIS-S) between April 2025 and January 2026, with exposure times ranging from 9--20.8 ks (PI: Reines, Proposal ID: 26700402). Each target galaxy has exactly one associated observation in this analysis (see Table \ref{tab:chandra_obs}).

\subsection{Data Reduction and Astrometry} \label{sec:astrometry}

Data were analyzed with the Chandra Interactive Analysis of Observations software package \citep[CIAO v4.17;][]{fruscione_ciao_2006}. First, all data were reprocessed using the $\texttt{chandra\_repro}$ method and the latest calibration files (CALB v4.12.0). After reprocessing our data to create new level 2 event files, we restrict our analysis to the S3 chip and generate an initial source list using \texttt{wavdetect}. We corrected $\textit{Chandra}$ astrometry by cross-matching with objects from the Sloan Digital Sky Survey Data Release 17 and our initial X-ray sources, excluding any source within $3''$ of the target galaxy coordinates supplied by GAMA. If there is at least one cross-matched source, we perform a simple translation correction. No correction was applied to IDs 3 and 8 due to lack of cross-matched sources. Astrometric corrections range from $0.''3-1''$.

To remove background flares, we created light curves of each observation while ignoring the initial source regions to isolate the background and used $\texttt{deflare}$ to find time periods with elevated background emission. Only $\sim260$s were removed from ID 8, and none of the other observations contain background flares. We report the effective exposure times with the \textit{Chandra} observations summary in Table \ref{tab:chandra_obs}.

Next, we rerun \texttt{wavdetect} while restricting the image to 0.5-7 keV. We use wavelet scales of 1.0, 1.4, 2.0, 2.8, and 4.0, with a point-spread function map of $39\%$ enclosed energy fraction at 2.3 keV. We set the threshold significance to $10^{-6}$, which corresponds to one false detection expected across the entire S3 chip.

Five of the seven target galaxies exhibit single X-ray point sources that appear spatially consistent with the optical galaxy centers reported by GAMA ($\lesssim 0.5''$ separation), with no additional nearby sources detected. At the mean redshift of our sample ($z = 0.16$), an angular separation of $0.5''$ corresponds to $\sim1.4$ kpc. The \texttt{wavdetect} algorithm does not identify X-ray sources in the vicinities of ID 1 or ID 5. However, as shown in Fig. \ref{fig:cutouts}, ID 1 exhibits some X-ray counts coincident with the galaxy center, while ID 5 shows no detectable X-ray emission. 

We investigate the positional uncertainties of the X-ray sources relative to their host galaxies using the 95\% positional uncertainties derived from Equation 12 of \citet{kim_chandra_2007}. In Fig. \ref{fig:xray_positions}, we overlay the X-ray source positions and their 95\% positional uncertainties on three-color Dark Energy Spectroscopic Instrument (DESI) images \citep{xu_desi_2023}, along with the optical centers reported by GAMA. The X-ray sources are located within $\lesssim0.5''$ of the optical centers of their host galaxies. For several sources, the statistical positional uncertainties from \citet{kim_chandra_2007} do not encompass the GAMA optical centers. However, these uncertainties characterize the X-ray source localization and do not account for residual uncertainties in the relative astrometric registration between the Chandra and optical reference frames. This is particularly relevant for IDs 3 and 8, for which no astrometric correction could be performed. IDs 1 and 10 have the largest positional uncertainties ($\sim1''$ and $\sim0.7''$, respectively). ID 1 was not independently detected by \texttt{wavdetect}, so its positional uncertainty was estimated at the GAMA position adopted for the source extraction and represents the expected localization uncertainty given its low source counts and off-axis angle. As seen in Fig. \ref{fig:cutouts}, the faint X-ray emission from ID 1 is concentrated near this optical position. ID 10 is independently detected, with its X-ray centroid lying close to the optical center and its positional uncertainty remaining within the central region of the host galaxy. Given the small observed offsets and the additional uncertainty associated with the relative astrometry, we consider the detected X-ray sources to be consistent with a nuclear origin.

For the remainder of this paper, we refer to galaxies and their associated X-ray sources interchangeably by their assigned IDs.

\begin{deluxetable}{ccccc}
\tabletypesize{\footnotesize}
\tablecaption{\label{tab:source_counts}Source and Background Counts}
\tablehead{
\colhead{ID} &
\multicolumn{2}{c}{0.5--2 keV} &
\multicolumn{2}{c}{2--7 keV} \\
\cmidrule(r){2-3}\cmidrule(l){4-5}
&
\colhead{Counts} &
\colhead{Background} &
\colhead{Counts} &
\colhead{Background}
}
\colnumbers
\startdata
1  & $<3$                  & 0.064 & $2^{+4}_{-2}$          & 0.157 \\
3  & $38^{+12}_{-10}$      & 0.036 & $31^{+11}_{-9}$       & 0.093 \\
5  & $<3$                  & 0.114 & $<3$                  & 0.136 \\
6  & $167^{+23}_{-21}$     & 0.100 & $171^{+23}_{-21}$     & 0.121 \\
8  & $97^{+18}_{-16}$      & 0.100 & $64^{+15}_{-13}$      & 0.050 \\
9  & $28^{+10}_{-8}$       & 0.043 & $44^{+13}_{-10}$      & 0.100 \\
10 & $<3$                  & 0.114 & $4^{+4}_{-3}$         & 0.143 \\
\enddata
\tablecomments{
(1) Source ID. (2) Observed source counts in the 0.5--2 keV band. (3) Expected background counts within the source aperture in the 0.5--2 keV band. (4) Observed source counts in the 2--7 keV band. (5) Expected background counts within the source aperture in the 2--7 keV band. For sources with $n\geq 10$ counts, uncertainties correspond to the 95\% confidence intervals of \citet{gehrels_confidence_1986}. For sources with $n<10$ counts, confidence intervals are estimated using the Bayesian formalism of \citet{kraft_determination_1991} when the source is detected at the 95\% confidence level given the expected background. Otherwise, we report the 95\% upper limit of 3 counts.
}
\end{deluxetable}

\begin{deluxetable*}{c c c c c c c c c c}
\tabletypesize{\footnotesize}
\tablecaption{\label{tab:xray_prop}X-ray Emission Properties} 
\tablehead{
\colhead{ID} & 
\multicolumn{3}{c}{Count Rate (counts ks$^{-1}$)} & 
\multicolumn{3}{c}{Flux ($10^{-15}$ erg s$^{-1}$ cm$^{-2}$)} & 
\multicolumn{3}{c}{$\log L_X$ (erg s$^{-1}$)} \\
\cmidrule(lr){2-4}\cmidrule(lr){5-7}\cmidrule(lr){8-10}
& \colhead{0.5--2 keV} & \colhead{0.5--8 keV} & \colhead{2--10 keV} 
& \colhead{0.5--2 keV} & \colhead{0.5--8 keV} & \colhead{2--10 keV} 
& \colhead{0.5--2 keV} & \colhead{0.5--8 keV} & \colhead{2--10 keV} 
}
\colnumbers
\startdata
1  &  $<0.333$ & $0.199^{+0.361}_{-0.196}$ & $<0.442$ & $<5.90$ & $3.69^{+6.35}_{-3.10}$ & $<9.88$ & $<41.49$ & $41.28^{+0.45}_{-0.80}$ & $<41.71$ \\
3  & $4.43^{+1.07}_{-1.27}$ & $8.20^{+1.90}_{-1.94}$ & $3.77^{+1.50}_{-1.20}$ & $83.3^{+29.7}_{-23.8}$ & $155^{+36}_{-37}$ & $80.2^{+31.8}_{-25.4}$ & $43.09^{+0.13}_{-0.15}$ & $43.36^{+0.09}_{-0.12}$ & $43.07^{+0.15}_{-0.16}$ \\
5 &  $<0.166$ & $<0.170$ & $<0.175$ & $<3.07$ & $<3.23$ & $<3.93$ & $<41.77$ & $<41.79$ & $<41.88$ \\
6  &  $10.3 \pm 1.6$ & $21.5 \pm 2.3$ & $11.3 \pm 1.7$ & $167 \pm 25$ & $398 \pm 43$ & $274 \pm 41$ & $42.66^{+0.06}_{-0.07}$ & $43.04^{+0.04}_{-0.05}$ & $42.88^{+0.06}_{-0.07}$ \\
8  & $17.6 \pm 3.5$ & $29.8 \pm 4.6$ & $12.4 \pm 3.0$ & $290 \pm 58$ & $542 \pm 84$ & $286 \pm 70$ & $43.57^{+0.08}_{-0.10}$ & $43.84 \pm 0.07$ & $43.57^{+0.09}_{-0.12}$ \\
9  &  $1.85^{+0.78}_{-0.61}$ & $4.86^{+1.12}_{-1.13}$ & $3.05^{+1.01}_{-0.82}$ & $97.2^{+40.8}_{-31.8}$ & $161 \pm 37$ & $77.8^{+25.5}_{-21.1}$ & $42.56^{+0.15}_{-0.18}$ & $42.78^{+0.08}_{-0.12}$ & $42.46^{+0.12}_{-0.14}$ \\
10  &  $<$0.239 & $0.242^{+0.305}_{-0.235}$ & $0.187^{+0.289}_{-0.183}$ & $<$3.96 & $4.53^{+5.77}_{-3.16}$ & $4.19^{+6.51}_{-3.31}$ & $<40.51$ & $40.53^{+0.35}_{-0.52}$ & $40.50^{+0.38}_{-0.68}$ \\
\enddata
\tablecomments{(1) Source ID. (2--4) Count rates in the 0.5--2 keV, 0.5--8 keV, and 2--10 keV bands. (5--7) Corresponding model unabsorbed fluxes. (8--10) X-ray luminosities. For IDs 3, 6, 8, and 9, fluxes are calculated using the best-fit spectral models described in \S\ref{sec:spectral_analysis}. For IDs 1, 5, and 10, we assume $\Gamma=2$ and Galactic absorption only. Uncertainties and upper confidence limits are quoted at the 95\% confidence level from \texttt{srcflux}.}
\end{deluxetable*}

{
\setlength{\tabcolsep}{3pt}
\begin{deluxetable}{cccccc}
\tabletypesize{\footnotesize}
\tablecaption{\label{tab:spectral_prop}X-ray Spectral Properties} 
\tablehead{
\colhead{ID} & 
\colhead{$\log n_{\rm H,Gal}$} & 
\colhead{$\log n_{\rm H,int}$} & 
\colhead{$\Gamma$} & 
\colhead{$\alpha_{\rm OX}$} &
\colhead{$\Delta\alpha_{\rm OX}$} \\
\colhead{} &
\colhead{(cm$^{-2}$)} &
\colhead{(cm$^{-2}$)} &
\colhead{} &
\colhead{} &
\colhead{}
}
\colnumbers
\startdata
1  & 20.19 & \ldots & \ldots & $<-1.2$ & $<-0.2$\\
3  & 19.96 & $<21.80$ & $2.16^{+0.88}_{-0.49}$ & $-1.3^{+0.2}_{-0.2}$ & $0.0^{+0.1}_{-0.1}$\\
5  & 20.40 & \ldots & \ldots & $<-1.3$ & $<-0.2$\\
6  & 20.51 & $<21.35$ & $1.75^{+0.32}_{-0.21}$ & $-1.1^{+0.2}_{-0.2}$ & $0.1^{+0.1}_{-0.1}$\\
8  & 20.08 & $<21.36$ & $1.89^{+0.38}_{-0.32}$ & $-1.2^{+0.2}_{-0.2}$ & $0.1^{+0.1}_{-0.1}$\\
9  & 20.68 & $22.03\;(<22.16)$ & $2.16^{+1.20}_{-1.13}$ & $-1.0^{+0.2}_{-0.2}$ & $0.1^{+0.1}_{-0.2}$\\
10 & 20.40 & \ldots & \ldots & $-1.5^{+0.2}_{-0.3}$ & $-0.5^{+0.2}_{-0.3}$
\enddata
\tablecomments{
(1) Source ID.
(2,3) Galactic and best-fit intrinsic hydrogen column densities, respectively. The lower 95\% confidence bounds on the intrinsic column density are unconstrained for all sources, so we report 95\% upper limits. For IDs 3, 6, and 8, the best-fit intrinsic column density is zero and only the upper limit is reported. For ID 9, the fit favors a nonzero intrinsic column density, so we report the best-fit value followed by the 95\% upper limit in parentheses.
(4) gives the best-fit photon index and 95\% confidence intervals for sources with sufficient counts.
Three-dot ellipses indicate that the source had too few counts to obtain a meaningful spectral fit.
(5) X-ray-to-UV spectral slope,
$\alpha_{\rm OX}=0.3838\log(l_{2\,\rm keV}/l_{2500})$,
where $l_{2\,\rm keV}$ and $l_{2500}$ are the rest-frame luminosity densities at 2 keV and 2500 \AA, respectively (see \S\ref{sec:weakness}).
For X-ray detected sources, quoted uncertainties are 95\% confidence intervals propagated from the X-ray and optical luminosity-density uncertainties.
For IDs 1 and 5, we instead report conservative upper limits obtained by combining the 95\% upper limits on $l_{2\,\rm keV}$ with the lower 95\% bounds on $l_{2500}$; the latter include both the broad H$\alpha$ measurement uncertainty and the 0.2 dex scatter in the broad H$\alpha$--$L_{5100}$ relation.
(6) Deviation from the expected value from the \citet{just_x-ray_2007} $\alpha_{\rm OX}$--$l_{2500}$ relation,
defined as $\Delta\alpha_{\rm OX}=\alpha_{\rm OX,obs}-\alpha_{\rm OX,exp}$.
For X-ray detected sources, quoted uncertainties are 95\% confidence intervals and account for the shared dependence of $\alpha_{\rm OX,obs}$ and $\alpha_{\rm OX,exp}$ on $l_{2500}$.
For IDs 1 and 5, conservative upper limits are calculated using the same procedure described for column (5).
More negative values indicate relatively weaker X-ray emission.}
\end{deluxetable}
}

\section{Analysis and Results}
\label{sec:analysis}

\subsection{X-ray Photometry} \label{sec:photometry}

We perform X-ray photometry to measure source counts and assess the significance of detections. Following the procedures presented in \citet{latimer_chandra_2021} and \citet{kimbrell_luminous_2024}, we extracted counts in a circular source aperture of radius $2''$, corresponding to the enclosed energy fraction of $90\%$ at 4.5 keV, centered on the \texttt{wavdetect} coordinates of the X-ray sources associated with our target galaxies. We used a background annulus centered on the source with inner and outer radii of 4 and $24''$, respectively. For IDs 1 and 5, which lack \texttt{wavdetect} coordinates, we centered the source and background regions on the galaxy coordinates taken from the GAMA survey. We verified that no X-ray sources lie within the background regions for any of the observations. Broad band images (0.5 -- 7 keV) of the target galaxies are shown in Fig. \ref{fig:cutouts}.

We consider a source to be detected if the source counts exceed the expected background counts in the source region at the 95\% confidence level using the Bayesian method of \citet{kraft_determination_1991} for Poisson-distributed data with low source counts. We checked each source in both the 0.5--2 keV and 2--7 keV ranges and find that six of the seven sources are detected in the 2--7 keV range, while IDs 1 and 10 are not detected in the 0.5--2 keV range. ID 5 is not detected in either band. We therefore use detection in the 2--7 keV band as our primary criterion for identifying an X-ray-detected source, while retaining ID 5 as a nondetection with upper limits in subsequent analyses.

The raw counts and $95\%$ confidence intervals in the soft (0.5 -- 2 keV) and hard (2 -- 7 keV) bands along with the expected number of background counts in the source region for each source are recorded in Table \ref{tab:source_counts}. In cases where counts are greater than or equal to ten, we assume that background counts are negligible and use the confidence intervals from \citet{gehrels_confidence_1986}. When the source counts are less than ten, we consider background counts and use the confidence intervals of \citet{kraft_determination_1991}.

We used \texttt{srcflux} to calculate count rates and model unabsorbed fluxes with 95\% confidence intervals in all relevant energy bands, with aperture corrections performed in \texttt{srcflux} using \texttt{psfmethod=arfcorr}. For sources with sufficient counts for spectral fitting (see \S\ref{sec:spectral_analysis}), we adopt the best-fit absorbed power-law model parameters, including the photon index ($\Gamma$) and intrinsic column density ($n_{\mathrm{H,int}}$). As described in \S\ref{sec:spectral_analysis}, Galactic absorption is fixed to the line-of-sight value from the HEASARC $n_{\mathrm{H}}$ tool using measurements from \citet{kalberla_leidenargentinebonn_2005}, while an additional absorption component at the source is allowed to account for intrinsic absorption.

For sources with insufficient counts to constrain spectral parameters (IDs 1, 5, and 10), we assume a typical AGN photon index of $\Gamma=2$ \citep{dong_x-ray_2012} and include only the fixed Galactic absorption. The resulting fluxes, luminosities, and upper limits therefore depend on this assumed spectral model and do not account for possible intrinsic absorption. If substantial intrinsic absorption is present, the intrinsic X-ray luminosities could be higher than inferred here.

We find the distances to the galaxies using \texttt{FlatLambdaCDM} from \texttt{astropy} and the spectroscopic redshifts reported by the GAMA survey. Using the unabsorbed fluxes described above and the calculated luminosity distances, we determine the luminosity of each source in each band. Fluxes and luminosities with their associated uncertainties in the 0.5--2 keV, 0.5--8 keV, and 2--10 keV bands are reported in Table \ref{tab:xray_prop}. For bands in which a source is not detected at the 95\% confidence level, we report 95\% upper limits on the flux and luminosity. Accordingly, for ID 1 we report upper limits in the 0.5--2 keV and 2--10 keV bands, while for ID 10 we report an upper limit in the 0.5--2 keV band. ID 5 is not detected in any of the three bands, and we therefore report upper limits in all three bands.

\begin{figure*}
\centering

\includegraphics[width=0.5\textwidth]{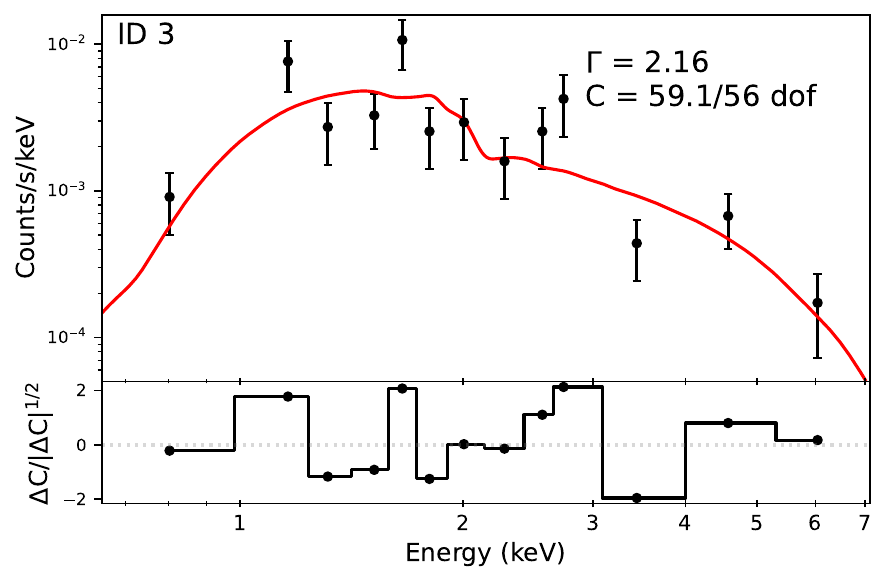}%
\hfill
\includegraphics[width=0.5\textwidth]{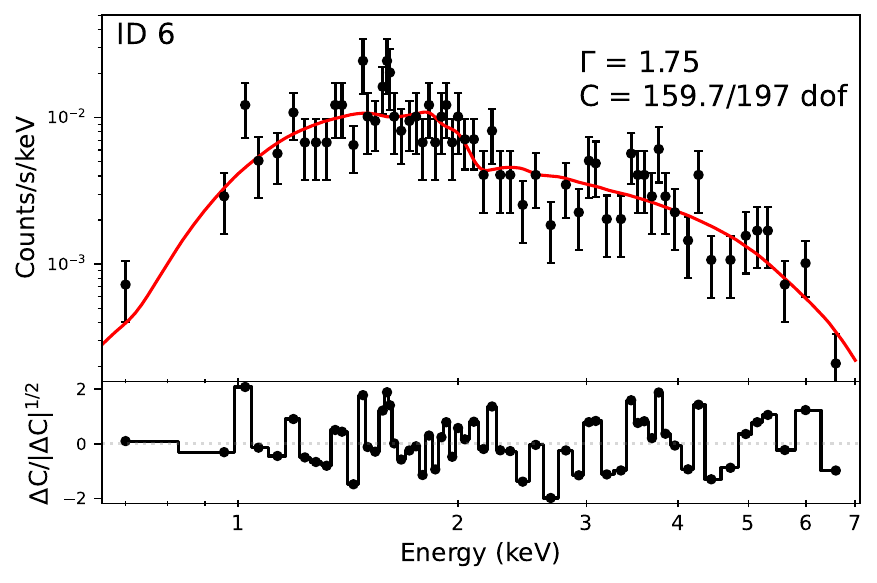}

\vspace{0.5em}

\includegraphics[width=0.5\textwidth]{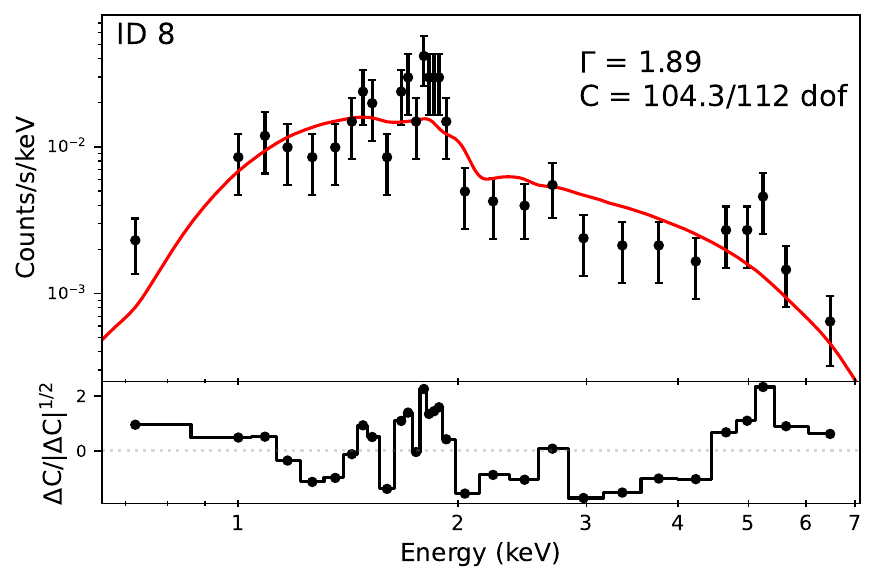}%
\hfill
\includegraphics[width=0.5\textwidth]{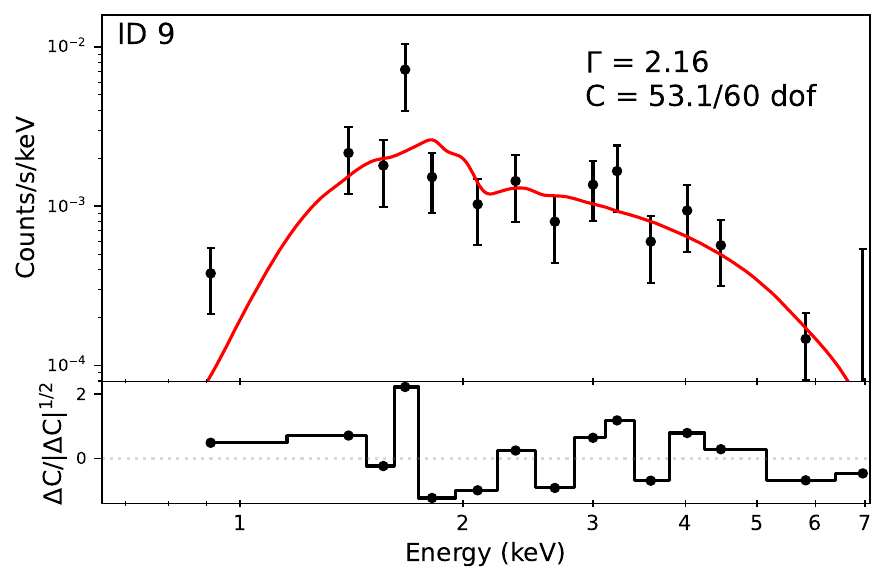}

\caption{X-ray spectra (black) and best-fit models (red) for IDs 3 (top left), 6 (top right), 8 (bottom left), and 9 (bottom right). The best-fit photon index ($\Gamma$) and Cash statistic (C) are shown in the top right corner of each panel. The spectra are modeled with a power law subject to fixed Galactic and variable intrinsic absorption (\texttt{xsphabs*xsphabs*powlaw1d} in Sherpa). For visualization, the spectra shown here are rebinned to $\geq5$ counts per bin, whereas the fits were performed using spectra grouped to $\geq1$ count per bin with Cash statistics. Residuals are shown below each spectral fit as $\Delta \mathrm{C}/|\Delta \mathrm{C}|^{1/2}$, where C is the Cash statistic \citep{cash_parameter_1979,kaastra_use_2017}. See \S\ref{sec:spectral_analysis} for additional details on the spectral fitting procedure and Table \ref{tab:spectral_prop} for the best-fit parameters.}
\label{fig:spectral_fits}
\end{figure*}

\subsection{Spectral Analysis} \label{sec:spectral_analysis}

We perform spectral analysis on four of the seven galaxies in our sample with sufficient counts. We model the extracted spectra over the $0.5$--$7$ keV range using an absorbed power-law model, \texttt{xsphabs*xsphabs*powlaw1d}, within the Sherpa v4.13 fitting package in CIAO. We include two absorption components: one fixed to the Galactic column density and a second allowed to vary freely to account for additional absorption intrinsic to the source. For IDs 3, 6, 8, and 9, we use the Cash statistic \citep{cash_parameter_1979, kaastra_use_2017} after re-binning the spectra to ensure at least one count per bin. We iteratively re-fit each spectrum using the best-fit parameters from the preceding fit as the initial values until the change in the fit statistic between consecutive fits satisfies $\Delta C < 10^{-4}$. The final best-fit models are shown in Fig. \ref{fig:spectral_fits}. The remaining observations (IDs 1, 5, and 10) have too few counts to obtain meaningful spectral constraints.

The resulting spectral properties are broadly consistent with unobscured or mildly obscured AGNs. The photon indices span $\Gamma \approx 1.7$--2.2, and most sources show no evidence for intrinsic absorption, with the exception of ID 9, which exhibits moderate intrinsic absorption. These properties are typical of AGN-powered X-ray emission. The best-fit parameters for each spectral model are recorded in Table \ref{tab:spectral_prop}.

\subsection{Black Hole Masses and Eddington Ratios} \label{sec:accretion}

A key property of our sample is that all galaxies have virial black hole mass estimates derived from broad Balmer line analysis \citep{salehirad_hundreds_2022}. These masses were calculated using the broad H$\alpha$ luminosity and FWHM with the relation in \citet{reines_dwarf_2013}, which is based on the approach in  \citet{greene_estimating_2005} and the size--luminosity relation of \citet{bentz_low-luminosity_2013}. For three galaxies in our sample (IDs 3, 5, and 8), Magellan spectroscopy yielded black hole mass estimates that differ by only $\sim0.15$ dex on average from the values reported by \citet{salehirad_hundreds_2022}, providing strong independent support for the broad-line measurements \citep{sturm_magellan_2025}.

\begin{figure}
    \centering
    \includegraphics[width=\linewidth]{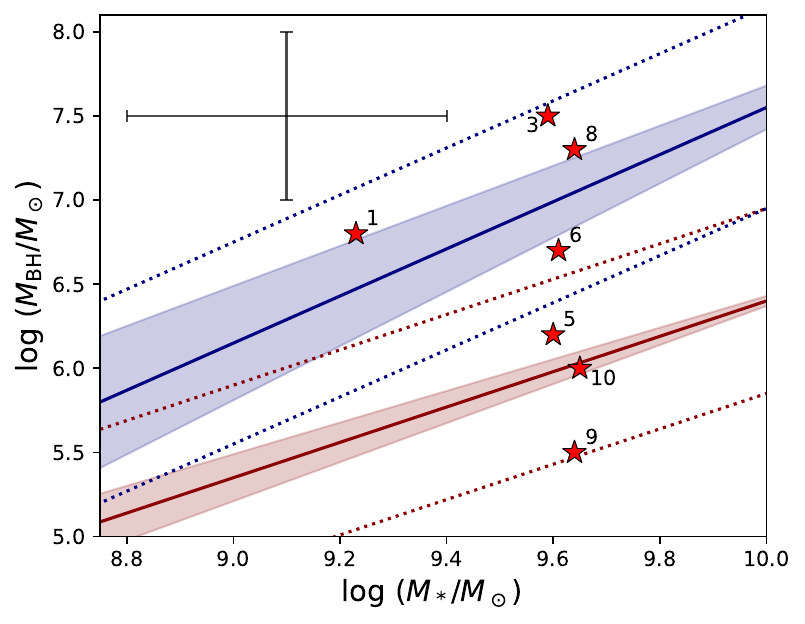}
    \caption{The solid red and blue lines denote the local $M_{BH}$ -- $M_{*}$ relation for AGN host galaxies and inactive galaxies from \citet{reines_relations_2015}, respectively. The shaded regions show the $1\sigma$ uncertainty for each fit, while the dotted lines illustrate the rms deviation of BH masses of 0.55 dex for the AGN relation and 0.60 dex for the inactive relation. The representative error bar in the upper left indicates typical uncertainties of 0.3 dex in stellar mass and 0.5 dex in virial black hole mass.}
    \label{fig:MBH_Mstar}
\end{figure}

More recently, \citet{cho_seoul_2023} presented an updated calibration that typically yields lower BH masses. We therefore recomputed black hole masses using the broad H$\alpha$ luminosities and FWHM values reported in Table 3 of \citet{salehirad_hundreds_2022} and Equation 6 of \citet{cho_seoul_2023}. The resulting masses are $\sim$40--95\% of the previously reported values. For comparison, we list both sets of masses and their corresponding Eddington ratios in Table \ref{tab:accretion_prop}. However, we adopt the masses reported by \citet{salehirad_hundreds_2022} as our fiducial values to remain consistent with the parent sample selection and with the black hole masses used to derive the AGN scaling relations of \citet{reines_relations_2015} in Fig. \ref{fig:MBH_Mstar} here.

Using these fiducial masses, we compare black hole mass and stellar mass in Fig. \ref{fig:MBH_Mstar}. Four of the seven galaxies (IDs 1, 3, 6, and 8) lie above the canonical AGN $M_{\rm BH}$--$M_*$ relation of \citet{reines_relations_2015}, appearing more consistent with the relation for inactive galaxies. This suggests that several galaxies in our sample may host comparatively massive black holes relative to their stellar masses. We note, however, that virial black hole mass estimates are indirect and carry systematic uncertainties of $\sim$0.5 dex \citep{vestergaard_determining_2006, shen_mass_2013}. We additionally illustrate a representative stellar mass uncertainty of 0.3 dex in Fig. \ref{fig:MBH_Mstar}. Given these uncertainties and the intrinsic scatter in the scaling relations, we do not draw strong conclusions from the precise location of individual sources, but note that the observed offsets are suggestive of diversity in black hole--host galaxy scaling relations at the low-mass end.

Examining the multiwavelength properties of the sample, we find no clear correspondence between position relative to the $M_{\rm BH}$--$M_*$ relation and AGN luminosity. IDs 3, 6, and 8 lie above the AGN relation and are also the most luminous sources in the sample across X-ray, mid-infrared, H$\alpha$, and [O \textsc{iii}] emission. In contrast, ID 1 also lies above the relation but is among the faintest sources, with weak X-ray emission, no detection in the 0.5--2 keV band, and relatively low optical and mid-infrared luminosities. Similarly, ID 9 lies below the AGN relation despite being luminous across multiple wavelengths, while IDs 5 and 10 lie closer to the AGN relation but are among the least luminous systems. These results suggest that galaxies both above and near local black hole--host galaxy scaling relations can exhibit a wide range of accretion states, highlighting the diversity of AGN properties in low-mass galaxies.

\begin{deluxetable*}{ccccccc}
\tabletypesize{\footnotesize}
\tablecaption{\label{tab:accretion_prop}Accretion Properties} 
\tablehead{
\colhead{ID} &
\colhead{$\log L_{\mathrm{H}\alpha,\mathrm{broad}}$ (erg s$^{-1}$)} &
\colhead{FWHM$_{\mathrm{H}\alpha,\mathrm{broad}}$ (km s$^{-1}$)} &
\colhead{$\log M_{\mathrm{BH}}$} &
\colhead{$\log \lambda_{\mathrm{Edd}}$} &
\colhead{$\log M_{\mathrm{BH}}$ (Cho+23)} &
\colhead{$\log \lambda_{\mathrm{Edd}}$ (Cho+23)}
}
\colnumbers
\startdata
1  & 40.94 & 2355 & 6.8 & $<-0.79$                      & 6.6 & $<-0.59$ \\
3  & 42.46 & 2206 & 7.5 & $-1.32\pm1.23$ & 7.5 & $-1.30\pm1.23$ \\
5  & 41.44 & 926  & 6.2 & $<-0.01$                      & 6.1 & $<0.09$ \\
6  & 41.37 & 1620 & 6.7 & $-0.72\pm1.22$                & 6.5 & $-0.56\pm1.22$ \\
8  & 42.63 & 1543 & 7.3 & $-0.61^{+1.22}_{-1.23}$       & 7.3 & $-0.57^{+1.22}_{-1.23}$ \\
9  & 40.80 & 574  & 5.5 & $0.05\pm1.23$  & 5.3 & $0.26\pm1.23$ \\
10 & 39.85 & 1582 & 6.0 & $-2.42^{+1.28}_{-1.40}$       & 5.6 & $-2.01^{+1.28}_{-1.40}$ \\
\enddata
\tablecomments{
(1) Galaxy ID.
(2) Broad H$\alpha$ luminosity from \citet{salehirad_hundreds_2022}.
(3) FWHM of broad H$\alpha$ from \citet{salehirad_hundreds_2022}.
(4) Black hole mass from \citet{salehirad_hundreds_2022} using the \citet{greene_estimating_2005} relation with the \citet{bentz_low-luminosity_2013} correction.
(5) Logarithm of the Eddington ratio using the \citet{salehirad_hundreds_2022} masses and bolometric luminosities obtained using the relation from \citet{duras_universal_2020}. For IDs 1 and 5, we report conservative upper limits derived from the 95\% upper limits on the X-ray luminosities together with the adopted scatter in the bolometric correction.
(6) Black hole mass using the updated calibration from \citet{cho_seoul_2023}.
(7) Logarithm of the Eddington ratio using the \citet{cho_seoul_2023} black hole masses and the same \citet{duras_universal_2020} bolometric luminosities. Eddington-ratio uncertainties include the X-ray luminosity uncertainty, the adopted scatter in the bolometric correction, and a 0.5 dex uncertainty in the virial black hole mass. All quoted uncertainties are reported at the 95\% confidence level.
}
\end{deluxetable*}

We estimate the bolometric luminosities ($L_{\rm bol}$) of our sources using the 2 -- 10 keV luminosities and the AGN bolometric correction ($K_X = L_{\rm bol}/L_X$) from \citet{duras_universal_2020}, who derive a luminosity-dependent relation between hard X-ray and bolometric luminosity:

\begin{equation} \label{eq:xray_bol}
    K_X (L_X) = a \Bigg[1 + \left(\frac{\log (L_X / L_{\odot})}{b}\right)^c \Bigg],
\end{equation}

\noindent where $L_X$ is the 2--10 keV luminosity, and $a = 15.33 \pm 0.06$, $b = 11.48 \pm 0.01$, $c = 16.20 \pm 0.16$, and the relation has a 0.37 dex scatter. We compute bolometric luminosities as $L_{\rm bol} = K_X L_X$, and derive Eddington ratios as $\lambda_{\rm Edd} = L_{\rm bol}/L_{\rm Edd}$, where $L_{\rm Edd} = 1.26 \times 10^{38} (M_{\rm BH}/M_\odot)\ \mathrm{erg\,s^{-1}}$.

For ID 5, which lacks an X-ray detection, and ID 1, which is undetected in the 2--10 keV band at the 95\% confidence level, we report conservative upper limits on the Eddington ratios based on the corresponding upper limits of X-ray luminosity. The derived black hole masses and Eddington ratios are listed in Table \ref{tab:accretion_prop}. The detected sources span a wide range of nominal Eddington ratios, $\log \lambda_{\rm Edd}\approx-2.42$--0.05. ID 9 has the highest nominal Eddington ratio, with a value near the Eddington limit, although the large propagated uncertainty prevents us from determining whether the source is accreting above or below this limit. Adopting the generally lower black hole masses derived using the calibration of \citet{cho_seoul_2023} shifts the inferred Eddington ratios upward, with nominal values of $\log \lambda_{\rm Edd}\approx-2.01$--0.26 for the detected sources.

These Eddington ratios are calculated using bolometric luminosities derived from the hard X-ray luminosity through the luminosity-dependent bolometric correction of \citet{duras_universal_2020}. Although the X-ray luminosities of our sample fall within the calibration range of the relation ($L_{\rm X}\gtrsim10^{40}$ erg s$^{-1}$), the calibration is based primarily on more luminous AGN. Furthermore, if some of our sources are intrinsically X-ray weak or X-ray strong relative to their optical emission, X-ray-based bolometric corrections may introduce systematic biases in the inferred accretion rates.

To assess the robustness of trends across our inferred accretion rates, we estimate bolometric luminosities using several commonly adopted bolometric corrections based on independent observables. Specifically, we compare our reported values from the hard X-ray correction of \citet{duras_universal_2020}, an optical continuum correction based on $L_{5100}$ from \citet{netzer_bolometric_2019}, an [O \textsc{iii}] based correction from \citet{heckman_present-day_2004}, and the luminosity-dependent hard X-ray correction of \citet{marconi_local_2004}. The resulting Eddington ratios are presented in Table \ref{tab:eddington_ratios}.

To provide an independent optical estimate, we first convert the broad H$\alpha$ luminosities reported by \citet{salehirad_hundreds_2022} to continuum luminosities at 5100 \AA\ using the empirical relation of \citet{greene_estimating_2005}, while adopting the
reported intrinsic scatter of $\sim0.2$ dex. We then apply the luminosity-dependent bolometric correction from Table 1 of
\citet{netzer_bolometric_2019},

\begin{equation}
\frac{L_{\rm Bol}}{L_{5100}} =
40 \left(\frac{L_{5100}}{10^{42}\ {\rm erg\ s^{-1}}}\right)^{-0.2}.
\end{equation}

\noindent Because \citet{netzer_bolometric_2019} does not provide a single statistical scatter for this relation, we adopt a conservative uncertainty of 0.35 dex, corresponding to the upper end of the uncertainty range illustrated for the $L_{5100}$ bolometric correction. We also note that the black hole masses of several sources in our sample fall below the mass range modeled by \citet{netzer_bolometric_2019}. We additionally estimate bolometric luminosities from the observed, uncorrected [O \textsc{iii}] luminosities using the relation of \citet{heckman_present-day_2004},

\begin{equation}
L_{\rm Bol} \approx 3500 \, L_{\rm [O III]},
\end{equation}

\noindent which has an intrinsic scatter of approximately 0.38 dex. We additionally note that broad [O \textsc{iii}] components were identified for IDs 3, 8, and 10 in the GAMA spectra \citep{salehirad_hundreds_2022}, and for IDs 3 and 5 in the Magellan spectra \citep{sturm_magellan_2025}, indicating possible ionized outflows. In these systems, the observed [O \textsc{iii}] luminosities may include contributions from outflowing gas, potentially introducing additional scatter into the inferred bolometric luminosities.

Finally, we apply the luminosity-dependent 2 -- 10 keV X-ray bolometric correction of \citet{marconi_local_2004}, which predicts bolometric corrections ranging from $L_{\rm Bol}/L_{2-10 \, {\rm keV}}\sim7$--25 across our sample. \citet{marconi_local_2004} estimate an uncertainty of approximately 0.1 dex in the logarithm of the X-ray bolometric correction. This provides a comparison with our reported \citet{duras_universal_2020} correction while relying on the same X-ray luminosity measurements.

\begin{deluxetable*}{ccccccccc}
\tabletypesize{\footnotesize}
\tablecaption{\label{tab:eddington_ratios}
Bolometric Luminosities and Eddington Ratios from Different Bolometric Corrections}

\tablehead{
\colhead{ID} &
\multicolumn{2}{c}{Duras+20} &
\multicolumn{2}{c}{Netzer 2019} &
\multicolumn{2}{c}{Heckman+04} &
\multicolumn{2}{c}{Marconi+04} \\
\cmidrule(lr){2-3}
\cmidrule(lr){4-5}
\cmidrule(lr){6-7}
\cmidrule(lr){8-9}
\colhead{} &
\colhead{$\log L_{\rm bol}$} &
\colhead{$\log\lambda_{\rm Edd}$} &
\colhead{$\log L_{\rm bol}$} &
\colhead{$\log\lambda_{\rm Edd}$} &
\colhead{$\log L_{\rm bol}$} &
\colhead{$\log\lambda_{\rm Edd}$} &
\colhead{$\log L_{\rm bol}$} &
\colhead{$\log\lambda_{\rm Edd}$} \\
\colhead{} &
\colhead{(erg s$^{-1}$)} &
\colhead{} &
\colhead{(erg s$^{-1}$)} &
\colhead{} &
\colhead{(erg s$^{-1}$)} &
\colhead{} &
\colhead{(erg s$^{-1}$)} &
\colhead{}
}
\colnumbers
\startdata
1  & $<43.62$       & $<-0.79$                 & $43.97\pm0.76$ & $-0.93\pm1.24$ & $44.25\pm0.75$ & $-0.66\pm1.23$ & $<42.89$ & $<-1.21$ \\
3  & $44.28\pm0.74$ & $-1.32^{+1.23}_{-1.23}$ & $45.02\pm0.75$ & $-0.58\pm1.24$ & $45.02\pm0.81$ & $-0.58\pm1.27$ & $44.34^{+0.27}_{-0.28}$ & $-1.26^{+1.02}_{-1.02}$ \\
5  & $<43.79$       & $<-0.01$                 & $44.32\pm0.76$ & $0.02\pm1.24$  & $44.55\pm0.75$ & $0.25\pm1.23$  & $<43.09$ & $<-0.41$ \\
6  & $44.08\pm0.73$ & $-0.72\pm1.22$           & $44.27\pm0.76$ & $-0.53\pm1.24$ & $44.08\pm0.83$ & $-0.72\pm1.28$ & $44.09\pm0.21$ & $-0.71\pm1.00$ \\
8  & $44.79\pm0.73$ & $-0.61^{+1.22}_{-1.23}$ & $45.14\pm0.76$ & $-0.26\pm1.24$ & $45.32\pm0.75$ & $-0.08\pm1.23$ & $44.96^{+0.23}_{-0.25}$ & $-0.44\pm1.01$ \\
9  & $43.65\pm0.74$ & $0.05\pm1.23$            & $43.87\pm0.76$ & $0.27\pm1.24$  & $44.26\pm0.75$ & $0.66\pm1.24$  & $43.58^{+0.25}_{-0.26}$ & $-0.02\pm1.01$ \\
10 & $41.68\pm0.82$ & $-2.42^{+1.28}_{-1.40}$ & $43.22\pm0.76$ & $-0.88\pm1.24$ & $44.02\pm0.75$ & $-0.08\pm1.23$ & $41.35^{+0.46}_{-0.73}$ & $-2.75^{+1.08}_{-1.22}$ \\
\enddata

\tablecomments{
(1) Galaxy ID.
(2,3) Bolometric luminosities and Eddington ratios derived using the
luminosity-dependent 2--10 keV X-ray bolometric correction of
\citet{duras_universal_2020}. For IDs 1 and 5, the reported values are
conservative upper limits obtained from the 95\% upper limits on the hard
X-ray luminosity together with the adopted scatter in the bolometric correction;
the Eddington-ratio limits additionally include the uncertainty in the virial
black hole mass.
(4,5) Bolometric luminosities and Eddington ratios derived from
$L_{5100}$ using the \citet{netzer_bolometric_2019} bolometric correction,
with $L_{5100}$ estimated from broad H$\alpha$ using
\citet{greene_estimating_2005}.
(6,7) Bolometric luminosities and Eddington ratios derived from the
uncorrected [O \textsc{iii}] luminosity using
\citet{heckman_present-day_2004},
$L_{\rm bol}=3500\,L_{\rm [O\,III]}$.
(8,9) Bolometric luminosities and Eddington ratios derived using the
luminosity-dependent 2--10 keV X-ray bolometric correction of
\citet{marconi_local_2004}. For IDs 1 and 5, the reported values are
conservative upper limits obtained from the 95\% upper limits on the hard
X-ray luminosity together with the adopted scatter in the bolometric correction;
the Eddington-ratio limits additionally include the uncertainty in the virial
black hole mass. Bolometric-luminosity uncertainties include the measurement uncertainty in the relevant luminosity together with the adopted uncertainty or scatter in the corresponding bolometric correction. Eddington-ratio uncertainties additionally include a 0.5 dex uncertainty in the single-epoch virial black hole masses. All quoted uncertainties are reported at the 95\% confidence level. }
\end{deluxetable*}

For each bolometric luminosity estimate, we propagate the measurement
uncertainty in the corresponding luminosity together with the intrinsic
scatter or adopted uncertainty in the bolometric correction. For the
Eddington ratios, we additionally adopt a 0.5 dex uncertainty in the virial black hole masses \citep{vestergaard_determining_2006,
shen_mass_2013}. All reported intervals correspond to the 95\% confidence
level. Because the uncertainties in the virial black hole masses and
bolometric corrections are large, the resulting Eddington-ratio intervals
span approximately an order of magnitude or more and should therefore be
interpreted as approximate indicators of accretion state rather than precise
measurements.

We find that the bolometric luminosities and Eddington ratios inferred from the optical tracers are generally larger than those inferred from the hard X-ray luminosities. This discrepancy is particularly notable for ID 5. The $L_{5100}$ and [O~\textsc{iii}] corrections yield nominal Eddington ratios near or above the Eddington limit, whereas the source is not detected in hard X-rays and the X-ray-based methods provide only conservative upper limits. Although the large propagated uncertainties prevent us from concluding that ID 5 is accreting at a super-Eddington rate, a systematic difference remains between the optical and X-ray-based estimates.

Overall, the different bolometric corrections agree that the sample spans a broad range of nominal accretion states, but the absolute Eddington ratios for individual sources are highly uncertain. IDs 8 and 9 generally occupy the high-accretion end of the sample, while ID 10 remains among the lowest-accretion sources in the X-ray-based estimates. However, the optical and X-ray prescriptions can differ substantially for individual objects, particularly IDs 5 and 10. These differences highlight the systematic uncertainty associated with applying empirical bolometric corrections to low-mass AGNs and motivate direct multiwavelength constraints on their spectral energy distributions.


\subsection{Expected X-ray Emission from Stellar Processes} \label{sec:stellar_emission}

Our target galaxies show multiwavelength evidence for hosting AGNs (see \S\ref{sec:observations}), and the X-ray luminosities of most detected sources further support an AGN origin (Table \ref{tab:xray_prop}). However, ID 10 has a relatively low X-ray luminosity that could plausibly arise from other processes, and ID 5 is an X-ray non-detection. We therefore estimate the possible contributions from X-ray binaries (XRBs) and ultraluminous X-ray sources (ULXs) in all our galaxies.

\vspace{.2cm}

\noindent
{\it XRBs:}

The expected X-ray luminosity from low-mass X-ray binaries (LMXBs) scales primarily with galaxy stellar mass \citep{gilfanov_low-mass_2004}, while the contribution from high-mass X-ray binaries (HMXBs) depends on the star-formation rate (SFR) and metallicity of the host galaxy \citep{grimm_high-mass_2003, lehmer_empirical_2024}. Our target galaxies have stellar masses in the range $\log (M_*/M_{\odot}) \sim 9.23-9.65$ (Table \ref{tab:chandra_obs}). Using the relation from \citet{gilfanov_low-mass_2004}, the expected LMXB contribution is $\log (L_X/{\rm erg~s^{-1}}) \lesssim 38.6$ across this mass range, well below the luminosities of our detected X-ray sources. We therefore focus on estimating the expected contribution from HMXBs, which requires estimates of the SFR and metallicity of each galaxy.

To calculate the SFRs of our target galaxies, we use data from the Wide-field Infrared Survey Explorer \citep[WISE;][]{wright_wide-field_2010}. Of the target galaxies, IDs 1 and 5 have 95\% upper limits in the W3 band rather than measurements with associated uncertainties. As such, the SFR estimates of IDs 1 and 5 are conservative upper limits. After subtracting the expected contribution of stellar processes using a fraction of the integrated W1 flux, we apply the relation from \citet{cluver_calibrating_2017} to estimate the galaxy-integrated SFR from $L_{12\mu{\rm m}}$:

\begin{equation}
\log \mathrm{SFR} =
(0.889 \pm 0.018)\,\log L_{12\mu m} - (7.76 \pm 0.15),
\end{equation}

\noindent where SFR is in $M_{\odot}\,\mathrm{yr}^{-1}$, $L_{12\mu m}$ is in $L_{\odot}$, and the relation has a $1\sigma$ scatter of 0.15 dex. The SFRs of our targets range from $\log(\mathrm{SFR}/M_{\odot}\,\mathrm{yr}^{-1})\sim-0.2$--1.5. AGN-heated dust may contribute to the W3 emission and therefore cause these SFRs, and the corresponding expected HMXB luminosities, to be overestimated. This is particularly relevant for IDs 3, 8, and 9, which fall within the mid-IR AGN selection region of \citet{jarrett_spitzer-wise_2011}.

Following \citet{latimer_chandra_2021}, we estimate the metallicities of our galaxies with the relation from \citet{pettini_oiiinii_2004}:

\begin{equation}
12 + \log(\text{O/H}) = 8.97 + 0.57 \times \log ([\text{N}\, \textsc{ii}]/\text{H}\alpha)
\end{equation}

\noindent

Using the emission-line fluxes from \citet{salehirad_hundreds_2022}, we find metallicities in the range $12+\log(\mathrm{O/H})\sim8.6$--8.74 for our targets. However, as discussed by \citet{latimer_chandra_2021}, this relation was calibrated for H\textsc{ii} regions and may overestimate the metallicity in galaxies hosting AGNs because AGN emission can enhance [N~\textsc{ii}]/H$\alpha$. This is particularly relevant for our sample because many of the galaxies exhibit elevated [N~\textsc{ii}]/H$\alpha$ ratios in the BPT diagram (Fig.~\ref{fig:NII_BPT}). The metallicities should therefore be treated with caution. We report the SFR and metallicity of each galaxy in Table \ref{tab:sfr_metallicity}.

\begin{figure}
    \centering
    \includegraphics[width=\linewidth]{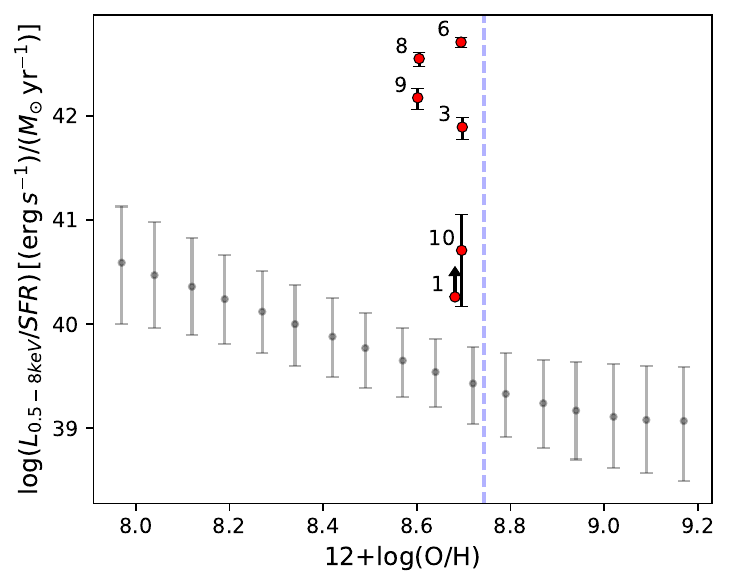}
    \caption{0.5--8 keV luminosities ($L_X$) of our X-ray sources and star-formation rates (SFRs) of their host galaxies compared with the empirical HMXB $L_X$--SFR--$Z$ relation from \citet{lehmer_empirical_2024}. The detected sources lie above the expected HMXB luminosities at their inferred SFRs and metallicities. The gray bars and points indicate the 16th, 50th, and 84th percentile HMXB luminosities over a range of metallicities. Error bars on the X-ray luminosities represent 95\% confidence intervals. For ID 1, the upward arrow and data point indicate a conservative lower limit on $L_X$/SFR obtained using the 95\% upper limit on the SFR and the 95\% lower limit on the X-ray luminosity. For ID 5, we show only the inferred metallicity (dashed blue line), because both $L_X$ and SFR are upper limits and their ratio is unconstrained.}
    \label{fig:LX_SFR_Z}
\end{figure}

We compare the 0.5--8 keV luminosities of our sources with the empirical $L_X$--SFR--$Z$ relation for HMXBs from \citet{lehmer_empirical_2024}. As shown in Fig.~\ref{fig:LX_SFR_Z}, the detected sources lie above the X-ray luminosities expected from HMXB populations at their inferred SFRs and metallicities. Because the metallicities inferred above may be overestimated, we consider the effect of lower metallicities when interpreting the possible HMXB contribution in \S\ref{sec:emission_origin}. ID 5 is not shown as an $L_X$/SFR measurement because both its X-ray luminosity and SFR are upper limits, leaving the ratio unconstrained.

\begin{deluxetable}{cccc}
\tabletypesize{\footnotesize}
\tablecaption{\label{tab:sfr_metallicity}Host-Galaxy Properties and Expected ULX Counts}
\tablehead{\colhead{ID} & \colhead{log SFR ($M_{\odot} \, \text{yr}^{-1}$)} & \colhead{12 + $\log(O/H)$} &  \colhead{Expected ULXs}}
\colnumbers
\startdata
1  &  $<0.22$ & 8.68 & $<0.12$ \\
3  &  1.50 & 8.70 & 2.05 \\
5 &  $<0.73$ & 8.74 & $<0.38$ \\
6 &  0.33 & 8.70 & 0.15 \\
8  &  1.30 & 8.60 & 1.37 \\
9  &  0.60 & 8.60 & 0.28 \\
10  &  -0.18 & 8.70 & 0.05 \\
\enddata
\tablecomments{(1) Galaxy ID. (2) Log star formation rate in solar masses per year. For IDs 1 and 5, the SFRs are 95\% upper limits derived from upper limits on their W3 luminosities. (3) Metallicity. (4) Expected number of ULXs with $L_X>10^{40}\,\mathrm{erg\,s^{-1}}$, adopting 0.07 ULXs per unit SFR for the metallicity range of our sample \citep{lehmer_metallicity_2021}. The expected ULX counts for IDs 1 and 5 are therefore also upper limits. Note that IDs 3, 8, and 9 have mid-IR colors indicating AGNs and therefore likely have artificially boosted mid-IR-derived SFRs and expected numbers of ULXs.}

\end{deluxetable}

\vspace{.2cm}

\noindent
{\it ULXs:}

In addition to the integrated emission from XRB populations, individual ultraluminous X-ray sources (ULXs) could potentially mimic low-luminosity AGNs. ULXs are typically associated with high-mass X-ray binaries undergoing super-Eddington accretion and are typically defined as off-nuclear X-ray sources with $L_X>10^{39}\,\mathrm{erg\,s^{-1}}$ \citep{kaaret_ultraluminous_2017}, but can reach luminosities of $L_X\gtrsim10^{40}\,\mathrm{erg\,s^{-1}}$, overlapping with the low-luminosity AGN regime.

Using the SFRs and metallicities derived above, we estimate the expected number of ULXs in each galaxy. From Fig.~6 of \citet{lehmer_metallicity_2021}, the expected number of ULXs with $L_X>10^{40}\,\mathrm{erg\,s^{-1}}$ is $\sim0.07$ per unit SFR over our metallicity range ($12+\log(\mathrm{O/H})\sim8.6$--8.7). Given our derived SFRs of $\log(\mathrm{SFR}/M_{\odot}\,\mathrm{yr}^{-1})\sim-0.2$--1.5, most galaxies are expected to host $\lesssim0.4$ such ULXs. IDs 3 and 8 have higher nominal expectations of $\sim2$ and $\sim1.4$, respectively, although these values are likely overestimated because their mid-IR-derived SFRs may contain significant AGN contributions. We report the expected number of ULXs for each galaxy in Table \ref{tab:sfr_metallicity}.

\begin{figure*}
\centering
\includegraphics[width=0.48\textwidth]{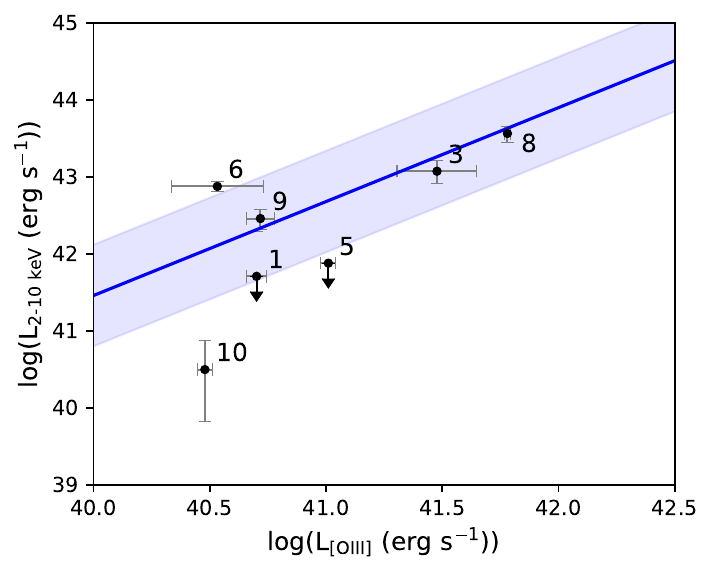}
\hfill
\includegraphics[width=0.48\textwidth]{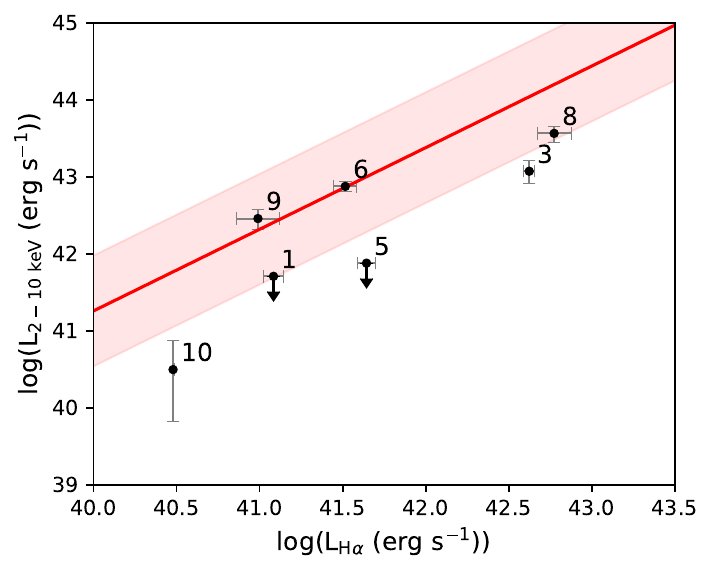}
\hfill
\includegraphics[width=0.48\textwidth]{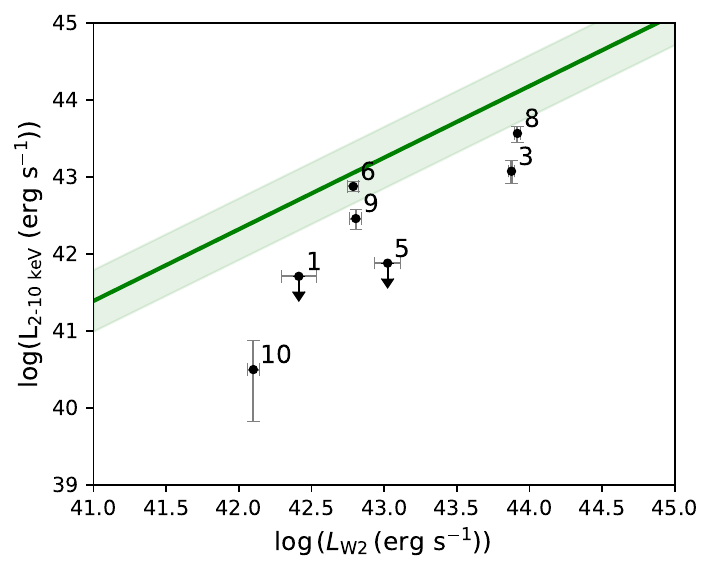}
\caption{Observed $L_{\mathrm{2-10\,keV}}$ versus $L_{\mathrm{[O\,\textsc{iii}]}}$ (\textit{upper left}) and $L_{\mathrm{H}\alpha}$ (\textit{upper right}) compared with the relations from \citet{panessa_x-ray_2006}, and $L_{\mathrm{2-10\,keV}}$ versus $L_{W2}$ (\textit{bottom}) compared with the relation from \citet{secrest_optically_2015}. The shaded regions represent the $1\sigma$ scatters of 0.66 dex, 0.72 dex \citep{latimer_chandra_2021}, and 0.4 dex for the [O~\textsc{iii}], H$\alpha$, and W2 relations, respectively. Error bars represent 95\% confidence intervals. Downward arrows for IDs 1 and 5 indicate 95\% upper limits on the 2--10 keV luminosities.}
\label{fig:xray_relations}
\end{figure*}

\subsection{Multiwavelength AGN Scaling Relations} \label{sec:scaling_results}

We compare the observed X-ray luminosities of our sources with empirical AGN scaling relations linking X-ray luminosity to H$\alpha$, [O~\textsc{iii}], and mid-infrared luminosities. In the top row of Fig.~\ref{fig:xray_relations}, we place our galaxies on the $L_{\mathrm{2-10\ keV}}$--$L_{\mathrm{H}\alpha}$ and $L_{\mathrm{2-10\ keV}}$--$L_{\mathrm{[O\,\textsc{iii}]}}$ relations from \citet{panessa_x-ray_2006}. These relations were derived using nuclear luminosities of Seyfert galaxies, and the H$\alpha$ relation includes both narrow and broad H$\alpha$ emission. We use the emission-line luminosities reported by \citet{salehirad_hundreds_2022} and adopt $1\sigma$ scatters of $\sim0.72$ dex and $\sim0.66$ dex for the H$\alpha$ and [O~\textsc{iii}] relations, respectively, as estimated by \citet{latimer_chandra_2021}.

We do not apply Balmer-decrement extinction corrections to the H$\alpha$ or [O~\textsc{iii}] luminosities. Because our targets exhibit broad Balmer emission, the measured H$\alpha$/H$\beta$ ratios may be affected by the decomposition of broad and narrow components, underlying stellar absorption, and differences in extinction between the broad- and narrow-line regions. We therefore adopt the published line luminosities without additional extinction corrections. The optical measurements may also contain contributions unrelated to the nuclear AGN emission. The GAMA spectra were obtained through a $2''$ fiber aperture, and star formation may contribute to the measured line luminosities, particularly for galaxies with significant star-forming contributions to their narrow emission lines. In addition, broad [O~\textsc{iii}] components have been identified in IDs 3, 5, 8, and 10 \citep{salehirad_hundreds_2022, sturm_magellan_2025}, such that outflowing gas may contribute to their [O~\textsc{iii}] luminosities. We therefore treat the optical line luminosities as approximate AGN tracers in these comparisons.

The X-ray-bright sources generally lie within the scatter of the published optical--X-ray scaling relations, while the lower-luminosity X-ray sources tend to occupy the lower envelope of the relations or fall below them. Because the optical luminosities have not been corrected for extinction, their intrinsic values may be larger than those shown in Fig.~\ref{fig:xray_relations}.

In the bottom panel of Fig.~\ref{fig:xray_relations}, we compare our sample with the $L_{\mathrm{2-10\ keV}}$--$L_{W2}$ relation from \citet{secrest_optically_2015}, which has an intrinsic scatter of $\sim0.4$ dex. All galaxies lie below the relation except ID 6, which is consistent within the scatter. IDs 1 and 10 lie more than an order of magnitude below the relation, while the upper limit for ID 5 also places it below the relation. The WISE W2 band has an angular resolution of $\sim6''.4$, comparable to the angular sizes of the galaxies in our sample, and therefore the measured W2 luminosities may include substantial host-galaxy emission in addition to emission from the AGN.

\subsection{X-ray Weakness} \label{sec:weakness}

To investigate whether our AGNs exhibit unusually weak X-ray emission relative to their optical luminosities, we follow the procedures described by previous studies and compute the X-ray-to-UV spectral slope, $\alpha_{\rm OX}$ \citep{dong_x-ray_2012, plotkin_x-ray_2016, paul_radio_2024}. This parameter is defined as

\begin{equation}
\alpha_{\rm OX} = 0.3838 \log \left( \frac{l_{2 \, \rm keV}}{l_{2500}} \right),
\end{equation}

\noindent where $l_{2\,\rm keV}$ and $l_{2500}$ are the unabsorbed monochromatic luminosity densities at rest-frame 2 keV and 2500 \AA, respectively.

We estimate $l_{2500}$ from the broad H$\alpha$ luminosity. We first convert the broad H$\alpha$ luminosity to the AGN continuum luminosity at 5100 \AA\ using the empirical relation of \citet{greene_estimating_2005}:

\begin{equation}
L_{\rm H\alpha} = 5.25 \times 10^{42}
\left(L_{5100}/10^{44}\ {\rm erg\ s}^{-1}\right)^{1.157}
{\rm erg\ s}^{-1}.
\end{equation}

Assuming an optical continuum of the form $f_{\nu}\propto\nu^{-0.44}$ \citep{vanden_berk_composite_2001}, we extrapolate the continuum luminosity to 2500 \AA\ and compute the corresponding luminosity density. We propagate the measurement uncertainty in the broad H$\alpha$ luminosity together with the 0.2 dex scatter in the broad H$\alpha$--$L_{5100}$ relation to estimate the uncertainty in $l_{2500}$.

For sources with sufficient counts for spectral fitting (IDs 3, 6, 8, and 9), we derive the 2 keV luminosity density from the best-fit X-ray spectral model by evaluating the unabsorbed power-law component at the observed-frame energy corresponding to rest-frame 2 keV. We estimate 95\% confidence intervals on the 2 keV flux density using the Sherpa \texttt{sample\_energy\_flux} routine, sampling the free spectral parameters from their covariance matrix 10,000 times while accounting for correlations between fitted parameters. For sources without sufficient counts for spectral fitting (IDs 1, 5, and 10), we use the Portable, Interactive Multi-Mission Simulator (PIMMS)\footnote{\url{https://heasarc.gsfc.nasa.gov/docs/software/tools/pimms.html}}, assuming $\Gamma=2$ and Galactic absorption, to convert the unabsorbed 2--10 keV fluxes determined in \S\ref{sec:photometry} to $l_{2\,\rm keV}$.

For the X-ray detected sources, the uncertainties reported in Table \ref{tab:spectral_prop} are 95\% confidence intervals propagated from the uncertainties in both $l_{2\,\rm keV}$ and $l_{2500}$. Because $l_{2500}$ enters both the observed and expected $\alpha_{\rm OX}$, we account for this shared dependence when propagating the uncertainty in $\Delta\alpha_{\rm OX}$. For IDs 1 and 5, which are not significantly detected in the hard band, we instead derive conservative upper limits by combining the 95\% upper limits on $l_{2\,\rm keV}$ with the lower 95\% bounds on $l_{2500}$.

We compare the resulting $\alpha_{\rm OX}$ values to the empirical $\alpha_{\rm OX}$--$l_{2500}$ relation from \citet{just_x-ray_2007},

\begin{equation}
\alpha_{\rm OX,exp} = -0.137 \, \log{l_{2500}} + 2.638,
\end{equation}

\noindent which has an intrinsic scatter of approximately 0.11 in $\alpha_{\rm OX}$ \citep{timlin_correlations_2020}. We define $\Delta\alpha_{\rm OX}=\alpha_{\rm OX}-\alpha_{\rm OX,exp}$, such that negative values indicate relatively weak X-ray emission.

Among the X-ray detected sources, we find $\alpha_{\rm OX}$ values spanning $-1.5$ to $-1.0$ and $\Delta\alpha_{\rm OX}$ values ranging from approximately $-0.5$ to $0.1$ (Table \ref{tab:spectral_prop}). IDs 3, 6, 8, and 9 are consistent with the expected relation within their 95\% confidence intervals. In contrast, ID 10 has $\Delta\alpha_{\rm OX}=-0.5^{+0.2}_{-0.3}$, indicating substantially weaker X-ray emission relative to its inferred UV luminosity. IDs 1 and 5 have conservative upper limits of $\Delta\alpha_{\rm OX}<-0.2$, providing evidence that they are also X-ray weak relative to their inferred UV luminosities, although the degree of their X-ray weakness remains unconstrained.

\section{Discussion} \label{sec:discussion}

\subsection{AGNs as the Origin of the X-ray Emission} \label{sec:emission_origin}

The X-ray luminosities of our detected sources generally favor accreting massive black holes over XRB populations as the origin of the observed emission. As shown in \S\ref{sec:stellar_emission}, the expected contribution from LMXBs is negligible, while most of our sources are substantially more X-ray luminous than expected from HMXBs given their host-galaxy SFRs and metallicities (Fig.~\ref{fig:LX_SFR_Z}). Because AGN emission may enhance [N~\textsc{ii}]/H$\alpha$, the true metallicities could be lower than those inferred from the \citet{pettini_oiiinii_2004} relation, which would increase the expected HMXB contribution. Even at lower metallicities, however, most of the detected sources remain more X-ray luminous than expected from HMXB populations alone. IDs 1 and 10 are less securely separated from the HMXB population, and HMXBs could plausibly contribute substantially to their observed X-ray emission if their true metallicities are sufficiently below our estimates. ID 5 is undetected in X-rays and therefore cannot be assessed with this comparison. Thus, while XRB emission could contribute substantially to the weaker detected sources, particularly IDs 1 and 10, it is unlikely to dominate the X-ray emission across the detected sample.

Individual ULXs could potentially produce the observed luminosities, but are also unlikely to explain the sample as a whole. Most of our galaxies are expected to host $\lesssim0.4$ ULXs with $L_X>10^{40}\,\mathrm{erg\,s^{-1}}$, although IDs 3 and 8 have larger nominal expectations of $\sim2$ and $\sim1.4$, respectively (Table~\ref{tab:sfr_metallicity}). These latter estimates are likely inflated because their mid-IR-derived SFRs may contain significant AGN contributions. More importantly, ULXs are generally off-nuclear sources, whereas the detected X-ray sources in our sample are spatially consistent with the optical galaxy nuclei when accounting for the relative X-ray--optical astrometric uncertainties (\S\ref{sec:astrometry}). The combination of the nuclear localization of the X-ray emission, the low expected ULX incidence in most hosts, and the X-ray luminosities in excess of the expected XRB populations therefore favors accreting massive black holes as the dominant origin of the detected X-ray emission.

\subsection{AGN Scaling Relations} \label{sec:scaling}

The broad agreement of the X-ray-bright sources with the optical--X-ray scaling relations suggests that their X-ray emission is broadly consistent with expectations for more massive AGNs. In contrast, the weaker X-ray sources tend to lie toward or below the lower envelope of these relations. Because we have not corrected the optical emission-line luminosities for extinction, their intrinsic H$\alpha$ and [O~\textsc{iii}] luminosities could be larger than those adopted here, which would increase rather than reduce the apparent X-ray deficits. Conversely, contributions from star formation or outflowing gas could increase the measured line luminosities, complicating the interpretation of these offsets.

The offsets are more pronounced relative to the mid-infrared relation, with most of the sample lying below the expected X-ray luminosity at a given W2 luminosity. However, the WISE measurements do not isolate the nucleus, and host-galaxy emission may contribute substantially to W2, particularly for the four galaxies that do not satisfy mid-infrared AGN color criteria. The mid-infrared offsets therefore cannot by themselves establish that these sources are intrinsically X-ray weak.

Taken together, the optical and mid-infrared comparisons suggest that some of the weaker X-ray sources may be underluminous relative to expectations from other AGN tracers, although systematic uncertainties prevent a definitive interpretation. The lack of strong intrinsic X-ray absorption in most sources, combined with the presence of broad-line emission, suggests that obscuration is unlikely to fully explain the observed offsets. In addition, scaling relations derived primarily from more massive AGNs may not apply directly to the low-mass systems studied here. We therefore examine the UV--X-ray spectral slope, $\alpha_{\rm OX}$, in the following section as an additional diagnostic of possible X-ray weakness.

\subsection{X-ray Weakness} \label{sec:xray_weakness}

Four of the seven sources are consistent with the expected $\alpha_{\rm OX}$--$l_{2500}$ relation, while IDs 1, 5, and 10 show evidence for relatively weak X-ray emission. We caution that $l_{2500}$ is inferred from broad H$\alpha$ rather than measured directly and may be affected by scatter in the broad H$\alpha$--continuum relation and internal extinction. In addition, the optical and X-ray observations are non-simultaneous, such that variability could contribute to the observed offsets. We therefore interpret these results as evidence for X-ray weakness relative to the inferred UV emission rather than definitive evidence that these sources are intrinsically X-ray weak.

ID 5 is particularly interesting because its possible X-ray weakness is also reflected in the discrepancy between its optical and X-ray based accretion rate estimates (\S\ref{sec:accretion}). The $L_{5100}$ and [O~\textsc{iii}] corrections yield nominal Eddington ratios near or above the Eddington limit, whereas the source is undetected in X-rays and the X-ray-based methods provide only upper limits. X-ray weakness has been observed in other high-Eddington-ratio, low-mass AGNs and may result from shielding of the compact X-ray-emitting region by a geometrically thick inner accretion disk \citep{paul_radio_2024}. Alternatively, X-ray variability or obscuration could suppress the observed X-ray emission relative to the non-simultaneous optical measurements \citep{ludlam_x-ray_2015}, while scatter in the optical bolometric corrections could cause the optical estimates to overpredict the bolometric luminosity. The current data do not allow us to distinguish among these possibilities.

Previous studies have similarly identified AGNs in low-mass galaxies that appear X-ray weak relative to their optical and UV emission \citep{baldassare_x-ray_2017, birchall_x-ray_2020, latimer_chandra_2021}, suggesting that departures from standard AGN scaling relations may occur in this regime. The range of $\alpha_{\rm OX}$ properties in our sample further demonstrates the diversity of X-ray emission among broad-line AGNs in low-mass galaxies \citep{paul_radio_2024}. Direct UV observations and deeper, multi-epoch X-ray observations will be necessary to determine whether the X-ray-weak sources are intrinsically weak and to investigate the roles of variability, absorption, and accretion state in producing the observed offsets.

\section{Conclusions} \label{sec:conclusion}

We have presented new \textit{Chandra} X-ray observations of seven broad-line AGNs in low-mass galaxies from the sample of \citet{salehirad_hundreds_2022}, with reported stellar masses $M_\star \lesssim 5 \times 10^9~M_\odot$. X-ray emission is detected in six of the seven galaxies, while ID 5 remains undetected. Five of the detected sources are independently identified as point sources and are spatially consistent with the optical galaxy nuclei after accounting for the relative X-ray--optical astrometric uncertainties; ID 1 instead shows faint X-ray emission coincident with the optical nucleus but is not independently detected by \texttt{wavdetect}.

The X-ray luminosities of most of the detected sources exceed those expected from stellar-mass X-ray binary populations at their inferred star-formation rates and metallicities. XRBs could contribute substantially to the weaker X-ray sources, particularly IDs 1 and 10, but are unlikely to dominate the detected sample as a whole. Individual ULXs are also unlikely to explain the sample given their expected incidence and the nuclear localization of the detected X-ray emission. Together, these results favor accretion onto central massive black holes as the dominant origin of the detected X-ray emission.

For the four sources with sufficient counts for spectral fitting, absorbed power-law models yield photon indices of $\Gamma\approx1.7$--2.2. Most show little evidence for intrinsic absorption beyond the Galactic foreground, while ID 9 exhibits moderate intrinsic absorption with $n_{\rm H,int}\sim10^{22}\,\mathrm{cm^{-2}}$.

Using virial black hole masses and several independent bolometric corrections, we find a broad range of inferred accretion rates across the sample. The hard-X-ray-based estimates yield nominal Eddington ratios spanning $\lambda_{\rm Edd}\sim0.004$--1.1, although the propagated uncertainties are substantial and the inferred values depend on the adopted bolometric correction. Bolometric luminosities and Eddington ratios inferred from the optical continuum and [O~\textsc{iii}] are generally larger than those inferred from the hard X-rays, indicating that the different tracers do not always provide consistent estimates of the accretion luminosity in these low-mass AGNs. This discrepancy is particularly notable for ID 5, for which the optical tracers yield nominal Eddington ratios near or above the Eddington limit, whereas the source is undetected in X-rays and the X-ray-based methods provide only upper limits. Given the large uncertainties, we cannot establish super-Eddington accretion in ID 5, but its discrepant optical and X-ray properties are consistent with the possibility of relative X-ray weakness.

Comparisons with established AGN scaling relations show mixed behavior. The brighter X-ray sources are broadly consistent with the optical $L_{\rm 2-10\,keV}$--H$\alpha$ and $L_{\rm 2-10\,keV}$--[O~\textsc{iii}] relations, while the weaker X-ray sources tend to lie toward or below their lower envelopes. Most of the sample also lies below the $L_{\rm 2-10\,keV}$--$L_{W2}$ relation, although host-galaxy contamination of the WISE photometry complicates the interpretation of these offsets. Our analysis of the UV-to-X-ray spectral slope, $\alpha_{\rm OX}$, provides additional evidence for relative X-ray weakness in a subset of the sample. IDs 3, 6, 8, and 9 are consistent with the expected relation between UV and X-ray emission, while ID 10 exhibits a substantial negative offset from the expected $\alpha_{\rm OX}$, and the hard-X-ray limits for IDs 1 and 5 imply $\Delta\alpha_{\rm OX}<-0.2$, where $\Delta\alpha_{\rm OX}$ measures the offset from the expected UV--X-ray relation. Because the UV luminosities are inferred from broad H$\alpha$ rather than measured directly, however, we cannot yet establish that these sources are intrinsically X-ray weak.

Overall, our results demonstrate a diversity of X-ray properties among broad-line AGNs in low-mass galaxies. While the stronger X-ray sources broadly resemble AGNs in more massive galaxies, some of the weaker sources show departures from standard multiwavelength scaling relations. Deeper and multi-epoch X-ray observations, direct UV measurements, and larger samples with virial black hole mass estimates will be important for determining the origin of these differences and for calibrating AGN scaling relations and bolometric corrections in the low-mass regime.

Finally, local low-mass AGNs provide important analogs for interpreting broad-line black hole candidates discovered by JWST at high redshift. JWST has revealed large numbers of high-redshift galaxies with broad H$\alpha$ emission \citep{harikane_jwstnirspec_2023, kocevski_hidden_2023, kokorev_uncover_2023, furtak_high_2024, greene_uncover_2024, ubler_ga-nifs_2024}, yet most remain weakly detected or undetected in X-rays \citep{matthee_little_2024, yue_stacking_2024, maiolino_jwst_2025, sacchi_chandra_2025}. Nearby low-mass galaxies therefore offer a valuable opportunity to study the X-ray properties and accretion physics of low-mass black holes in a regime that remains difficult to access directly at early cosmic times.

\section*{Acknowledgments}

The authors thank the anonymous referee for their helpful comments that improved the manuscript. Support for this work was provided by NASA through Chandra Award Number GO5-26083X issued by the Chandra X-ray Observatory Center, which is operated by the Smithsonian Astrophysical Observatory for and on behalf of NASA under contract NAS8-03060. A.E.R. gratefully acknowledges support for this work provided by the NSF through CAREER award 2235277.

\software{Astropy \citep{astropy_collaboration_astropy_2013,astropy_collaboration_astropy_2018,astropy_collaboration_astropy_2022}}

\bibliography{chantex}
\bibliographystyle{aasjournal}

\end{document}